\documentclass[pdflatex,sn-basic]{sn-jnl}% Basic Springer Nature Reference Style/Chemistry Reference Style

\usepackage{xcolor,colortbl}
\newcolumntype{s}{>{\columncolor{lightgray}}c}

\jyear{2021}%

\theoremstyle{thmstyleone}%
\theoremstyle{thmstyletwo}%

\theoremstyle{thmstylethree}%

\usepackage{etoolbox}   
\makeatletter
\patchcmd{\@maketitle}{\artauthors}{{\artauthors}}{}{}
\makeatother
\begin{document}

\title[Justifying the Volatility of S\&P 500 Daily Returns]{Justifying the Volatility of S\&P 500 Daily Returns}%A Justification for the Gradual Increase in Volatility of Daily Returns of the S\&P 500 Index

%%=============================================================%%
%% Prefix	-> \pfx{Dr}
%% GivenName	-> \fnm{Joergen W.}
%% Particle	-> \spfx{van der} -> surname prefix
%% FamilyName	-> \sur{Ploeg}
%% Suffix	-> \sfx{IV}
%% NatureName	-> \tanm{Poet Laureate} -> Title after name
%% Degrees	-> \dgr{MSc, PhD}
%% \author*[1,2]{\pfx{Dr} \fnm{Joergen W.} \spfx{van der} \sur{Ploeg} \sfx{IV} \tanm{Poet Laureate} 
%%                 \dgr{MSc, PhD}}\email{iauthor@gmail.com}
%%=============================================================%%

\author{\fnm{\quad\quad\ \  Author: Hayden} \sur{Brown}, ORCID: 0000-0002-2975-2711}\email{haydenb550222@gmail.com}%0000-0002-2975-2711

%\affil[1]{\orgdiv{Department of Mathematics and Statistics}, \orgname{University of Nevada, Reno}, \orgaddress{1664 \street{N. Virginia Street}, \city{Reno}, \postcode{89557}, \state{Nevada}, \country{USA}}}

%%==================================%%
%% sample for unstructured abstract %%
%%==================================%%

\abstract{Over the past 60 years, there has been a gradual increase in the volatility of daily returns for the S\&P 500 Index. Hypothetically, suppose that market forces determine daily volatility such that a daily leveraged S\&P 500 fund cannot outperform a standard S\&P 500 fund in the long run. Then this hypothetical volatility happens to support the increase in volatility seen in the S\&P 500 index. On this basis, it appears that the classic argument of the market portfolio being unbeatable in the long run is determining the volatility of S\&P 500 daily returns. Moreover, it follows that the long-term volatility of the daily returns for the S\&P 500 Index should continue to increase until passing a particular threshold. If, on the other hand, this hypothesis about market forces increasing volatility is invalid, then there is room for daily leveraged S\&P 500 funds to outperform their unleveraged counterparts in the long run.}%When the initial investment is in the S\&P Composite Index and $2\leq k\leq 16$, then the initial investment must be at least $k$ times the amount of each withdrawal. 
%Suppose that between rebalancing times, the assets have independent log-returns.

%Consider a pooled annuity fund investing in $n$ assets with discrete-time rebalancing. At time 0, each annuitant makes an initial contribution to the fund, committing to a predetermined schedule of withdrawals. Require the fund to be closed in the sense that no additional individuals can join the fund after time 0, and no withdrawals can be made outside of the predetermined schedule. Further require annuitants to be homogeneous in the sense that their initial contributions and predetermined withdrawal schedules are identical, and their mortality distributions are identical and independent. 


%\acknowledgments{I am grateful to Andrey Sarantsev for helpful remarks about early drafts of this manuscript.}

%\pacs[JEL Classification]{C22, E27, G11}

%\pacs[MSC Classification]{60E15, 60J70, 91B70}

%\pacs[Acknowledgments]{}

\maketitle

\section{Introduction}\label{sec1}
The return realized after buying and holding a daily leveraged exchange traded fund (ETF) for more than one day is largely dependent on the mean and volatility of the daily log-returns of the underlying index. When the mean is positive, increased volatility generally leads to decreased return. This is a result of the compounding effect of daily leveraged returns, which are particularly sensitive to changes in volatility. If volatility is sufficiently low, then a daily leveraged ETF will outperform its underlying (unlevered) index. However, there is the classic idea that the market portfolio should be unbeatable in the long-run. In this vein, a daily leveraged S\&P 500 ETF should not be able to beat an ETF tracking the S\&P 500 Index in the long-run. The main goal of this work is to determine what level of volatility leads to a standard S\&P 500 ETF dominating its daily leveraged counterparts in the long run.%But what level of daily volatility will guarantee this dominance of the standard S\&P 500 ETF?  

\subsection{Literature Review}
There are many ways to define and model volatility. For example, see \cite{avramov2006impact}, \cite{engle2008economic}, \cite{mcmillan2000forecasting} or \cite{takahashi2021forecasting}. The main measure of volatility used here has the flavor of realized volatility. In particular, volatility is taken to be the average squared daily percentage change over a specified number of consecutive trading days. %This measure is dubbed \textit{daily volatility}.

Most of the literature deals with short-term volatility; volatility is modeled as a stochastic process that evolves continuously in time or discretely over short-term time steps, like daily, weekly or monthly. Here, the goal is not to focus on short-term fluctuations in volatility. Rather, the goal is to examine how volatility has behaved over the past 60 years and then gain a reasonable idea of what the long-term (10+ years) volatility could be like in the future. From there, it is possible to estimate the future performance of a daily leveraged ETF relative to an ETF tracking the underlying index.

%\cite{mcmillan2000forecasting}
%\cite{engle2008economic} Realized volatility $r_i^2$
%\cite{avramov2006impact,takahashi2021forecasting}

With respect to the market portfolio containing all stocks in the major US stock exchanges, it is clear that volatility in daily returns has increased over time, but not volatility in monthly returns \citep{washer2016increasing}. This apparent increase in the volatility of daily returns serves as motivation for the examination presented here. Given that increased volatility tends to reduce daily leveraged ETF returns, it is worth pinpointing the connection between daily leveraged ETF returns and volatility in daily returns.

There is some empirical evidence showing that ETFs increase the volatility of their underlying stocks' prices \citep{ben2018etfs}. In particular, a stock that is owned by an ETF generally has a higher volatility compared to a stock that is not owned by an ETF. Moreover, increased ownership of a stock by ETFs also leads to increased volatility. The higher volatility appears to be a result of arbitrage activity between an ETF and its underlying stocks. Still, it is unclear exactly how much of the increase in volatility can be attributed to ETF prevalence versus other factors. Regardless, volatility has increased in the past, and it appears reasonable for further increases to occur in the future. Thus, it is important to investigate how future increases could affect daily leveraged ETF returns.

In \cite{conrad2015anticipating}, volatility of daily S\&P 500 log-returns is modeled using macroeconomic variables; the focus is on forecasts 1, 126 and 252 days into the future. In that setting, there is strong evidence supporting the use of macroeconomic variables as predictors of volatility. In contrast, the focus here is on the long-term (10+ years) volatility of daily S\&P 500 returns.

%In \cite{conrad2015anticipating}, the mid-term (126 and 252 days) volatility of daily S\&P 500 log-returns is modeled using macroeconomic variables. In that setting, there is strong evidence supporting the use of macroeconomic variables as predictors of mid-term volatility. In contrast, the focus here is on the long-term (multiple decades) volatility of daily S\&P 500 returns. %In contrast, the goal here is to determine the long-term (multiple decades) volatility of daily S\&P 500 returns. 

Using a model that incorporates memory and adapts to level changes, \cite{perron2010long} predict the volatility of daily S\&P 500 returns. Being adaptive and data driven, that model does not address the fundamental cause of a level change. Here, a goal is to identify level changes in the long-term volatility of daily S\&P 500 returns. Moreover, it is suspected that the apparent upward trend in these level changes may have something to do with market calibration of the relative performance between daily leveraged S\&P 500 ETFs and ETFs tracking the S\&P 500 index. In this context, there may be sub-level changes happening in the short-term, but those are not of interest here.

A connection between volatility in daily returns and long-term daily leveraged ETF returns has been made in \cite{brown}. In particular, a theoretical link is given that bases the performance of a daily leveraged ETF relative to its underlying index on the mean and volatility of daily log-returns of the underlying index. That connection is expanded upon here, with the goal of determining what level of volatility prevents a daily leveraged ETF from outperforming an ETF tracking its underlying index.%precludes outperformance of a leveraged ETF. 

In order to verify the generality of results beyond just S\&P 500 ETFs, linear programming is used to bound the error on estimations of the relative performance between a daily leveraged ETF and its unlevered counterpart; see \cite{winston2004operations} and \cite{klotz2013practical} for a review about linear programs (LPs). The bounding method relies on LPs that optimize the expectation of a particular function of a discrete random variable, subject to some incomplete information about the random variable's distribution. Note that the more general version of this optimization problem over all measurable random variables can be approached with semi-infinite linear programming (for example, see \cite{prekopa2016relationship}, \cite{mehrotra2014models} and \cite{goberna2002linear}). However, the added complications associated with considering all measurable random variables is not necessary to generate the bounds.

  %semiinfinite LP review update \cite{lopez2007semi}

\subsection{Summary of Results}
Bounds are given on the mean of squared daily percentage changes of the underlying index that indicate when a daily leveraged ETF will (approximately) outperform or underperform an ETF tracking its underlying index. The bounds are a function of the mean daily log-return of the underlying index and the ETFs' fees. They are based on an approximation of daily log-returns that is a quadratic function of the underlying index's daily percentage changes. This approximation is shown to hold well for the S\&P 500 Index and some foreign indexes, especially over 10+ year periods. 

Applications reveal that despite the increase in volatility of daily returns already seen in the S\&P 500, there may still be a noticeable additional increase to come, assuming that market forces are determining volatility so that a daily leveraged ETF cannot beat an ETF tracking its underlying index in the long run. If this is not the case, then there is a potential opening for daily leveraged ETFs to dominate. 

\subsection{Organization}
Section \ref{s:notation} sets up the notation and describes the data. Results are given in section \ref{s:theory} with step-by-step derivations. Closing remarks and a discussion of related future research ideas are given in section \ref{s:conclusion}.

\section{Preliminaries}\label{s:notation}
For consistency, the notation from \cite{brown} is reproduced and used here. Let $C_i$ denote the adjusted closing price of trading day $i$ for a particular stock market index I. Then $\{C_i\}_{i=0}^n$ is a sequence of adjusted closing prices for $n+1$ consecutive trading days. Note that adjusted closing prices account for dividends and stock splits, but not inflation. 

Let $X_i=C_{i}/C_{i-1}-1$ for $i=1,...,n$. Then $\{100\cdot X_i\}_{i=1}^n$ is the sequence of $n$ percentage changes between adjusted closing prices. Observe that 
\begin{equation*}
\prod_{i=1}^n(1+X_i)=\frac{C_n}{C_0}.
\end{equation*}

Denote the daily leveraged version of I as LxI, where L indicates the amount of leverage. For example, 3xI indicates the index tracking I with 3x daily leverage. The adjusted closing prices of LxI are given by 
\begin{equation*}
C_i^L:=C_0\cdot\prod_{k=1}^i(1+LX_k), \quad i=0,...,n.
\end{equation*}
So the log-returns realized by going long in LxI from the close of trading day $0$ to the close of trading day $n$ are given by
\begin{equation*}
\log \frac{C_n^L}{C_0}=\sum_{i=1}^n\log(1+LX_i).
\end{equation*}
Note that here, $\log$ refers to the natural logarithm. 

Denote the ETF version of LxI as LxI$_r$, where $r$ is the annual expense ratio, compounded on a daily basis. In subsequent text, the word \textit{index} refers to an ETF having $r=0$, and vice versa. Assuming 252 trading days in a year, the log-return of LxI$_r$ after $n$ days is given by
\begin{equation}\label{eq:RnrL}
R_{n,r}^L:=\log\frac{C_n^L}{C_0}+n\log\Big(1-\frac{r}{252}\Big).
\end{equation}
Let $d(L)$ denote the annualized $R_{n,0}^{L}-R_{n,0}^1$, i.e. $(R_{n,0}^{L}-R_{n,0}^1)\cdot 252/n$. Dividing $R_{n,0}^{L}-R_{n,0}^1$ by $n$ provides an average daily return difference, and multiplying that number by 252 yields an annualized return difference that is easier to interpret. Also let
\begin{equation}\label{eq:uv}
u=\frac{1}{n}\cdot R_{n,0}^1\ ,\quad v=\frac{1}{n}\sum_{i=1}^nX_i^2.
\end{equation}

From here on, let I refer to the S\&P 500 index. The expense ratio for SPY, a very popular ETF tracking the S\&P 500 index is .0945\% ($r=.000945$). The expense ratio for most leveraged S\&P 500 ETFs is .95\% ($r=.0095$). Companies offering popular leveraged S\&P 500 ETFs include Direxion and ProShares. %Let $r_0=.000945$ and $r_1=.0095$.

For quick reference, table \ref{t:fv} shows important functions and variables that have been or will be introduced.

\begin{table}
\begin{center}
\caption{Important functions and variables}\label{t:fv}
{\renewcommand{\arraystretch}{1}
\begin{tabular}{ ll } 
\toprule
\textbf{Function or variable} & \textbf{Description} \\
 \toprule
 $u$ & \begin{tabular}{@{}l@{}}average log-return of the unleveraged index over $n$ \\consecutive trading days\end{tabular}\\[2.5ex]
 $v,\ m_3,\ m_4$ & \begin{tabular}{@{}l@{}l@{}}average second, third and fourth power of the daily \\percentage change for the unleveraged index over $n$ \\consecutive trading days\end{tabular}\\[4ex]
$R_{n,r}^{L}$ & \begin{tabular}{@{}l@{}}log-return of a daily leveraged ETF after $n$ consecutive\\trading days (leverage multiple is $L$ and fee is $r$)\end{tabular}\\[2.5ex]
$d(L)$& annualized $R_{n,0}^{L}-R_{n,0}^1$, given by $(R_{n,0}^{L}-R_{n,0}^1)\cdot 252/n$\\[1ex]
 $n\cdot g(L)$ & approximates $R_{n,0}^{L}-R_{n,0}^1$, given by $n(L-1)(u-Lv/2)$\\ [1ex]
 $\widetilde{g}(L)$ & given by $m_3(L^3-L)/3-m_4(L^4-L)/4$\\[1ex]
 $n\cdot [g(L)+\widetilde{g}(L)]$ & approximates $R_{n,0}^{L}-R_{n,0}^1$\\[1ex]
 $L^*$ & argmax (with respect to $L$) of $R_{n,0}^{L}-R_{n,0}^1$\\[1ex]
 $\hat{L^*}$ &  argmax of $g(L)$, given by $u/v+1/2$, approximates $L^*$\\ [1ex]
 $\tilde{L^*}$ &  argmax of $g(L)+\widetilde{g}(L)$, approximates $L^*$\\ 
 \toprule
\end{tabular}}
\end{center}
\end{table}

\subsection{Data}\label{s:data}
\begin{figure}
  \includegraphics[width=\linewidth]{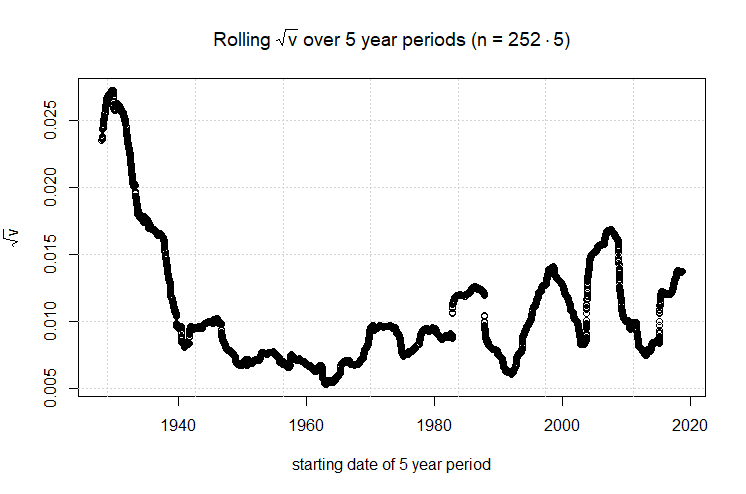}
  \caption{Recall \eqref{eq:uv}. $\sqrt{v}$ measures the average volatility in daily returns of the underlying index.}
  \label{fig:rollingv}
\end{figure}

Adjusted closing prices of the S\&P 500 Index are taken from \url{https://finance.yahoo.com}, spanning December 29, 1927 to September 29, 2023. Figure \ref{fig:rollingv} shows how $\sqrt{v}$ has been increasing over time. After 1960, there is an increasing trend in $\sqrt{v}$. It is impossible to say with certainty whether this trend will continue into the future. However, the advancement of this trend into the future appears plausible.

Adjusted closing prices of the Hang Seng Composite Index, DAX 30 Index and the CAC 40 Index are taken from \url{https://www.macrotrends.net}. Data used spans November 26, 1990 to November 14, 2023. These three foreign indexes track the performance of the largest public companies in Hong Kong, Germany and France, respectively. 

\begin{figure}
  \includegraphics[width=\linewidth]{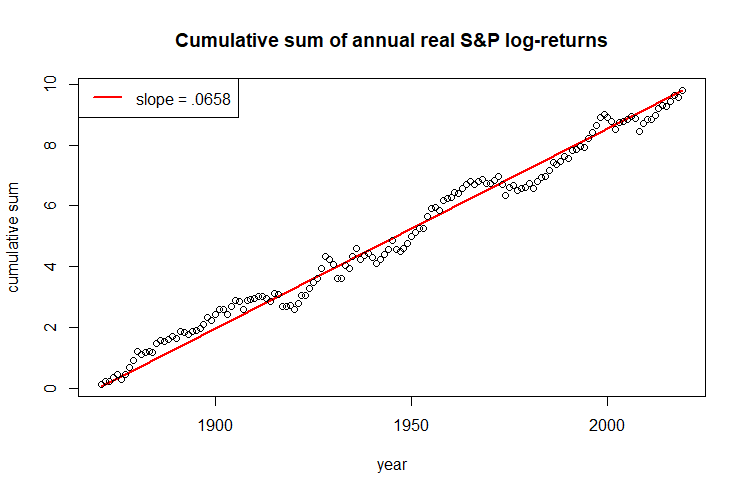}
  \caption{The cumulative sum of S\&P Composite Index annual real log-returns is illustrated over 150 years. The slope of the linear trend is equal to the mean annual real log-return over that period.}
  \label{fig:sptrend}
\end{figure}

Annual real returns of the S\&P Composite Index from 1871 to 2020 are taken from \url{http://www.econ.yale.edu/~shiller/data.html}. The S\&P 500 only goes back to 1957, so Cowles and Associates (1871 to 1926) and the Standard \& Poor 90 (1926 to 1957) are used as backward extensions. Relevant variables from the data are described in table \ref{t:data}. %, collected for easy access at \url{https://github.com/HaydenBrown/Investing}.
\begin{table}
\begin{center}
\caption{Data variable descriptions}\label{t:data}
\begin{tabular}{ ll } 
\toprule
\textbf{Notation} & \textbf{Description} \\
 \toprule
 $P$ & average monthly close of the S\&P composite index \\ 
 $D$ & dividend per share of the S\&P composite index \\ 
 $J$ & January consumer price index \\ 
 \toprule
\end{tabular}
\end{center}
\end{table}
Inflation and dividend adjusted (i.e. real) annual returns are computed using the consumer price index, the S\&P Composite Index price and the S\&P Composite Index dividend. Use the subscript $k$ to denote the $k$th year of $J$, $P$ and $D$. Then the real return for year $k$ is given by $((P_{k+1}+D_k)/P_k)\cdot(J_k/J_{k+1})$. Figure \ref{fig:sptrend} illustrates the remarkable stability of the S\&P Composite Index real returns over the past 150 years. Supposing this stability continues, the long-term average annual log-return of the S\&P 500 should be around .0658 plus the long-term average annual log-return of the consumer price index. Some possibilities are collected in table \ref{t:posrr}.
\begin{table}
\begin{center}
\caption{Long-term average annual log-return for different inflation rates.}\label{t:posrr}
\begin{tabular}{ ll } 
\toprule
\textbf{Long-term annual inflation} & \textbf{Long-term annual log-return} \\
 \toprule
 0\% & .0658 \\ 
 1\% & .0757 \\ 
 2\% & .0856 \\ 
 3\% & .0953 \\
 4\% & .1050 \\
 \toprule
\end{tabular}
\end{center}
\end{table}

%<a href='https://www.macrotrends.net/2594/hang-seng-composite-index-historical-chart-data'>Hang Seng Composite Index - 30 Year Historical Chart</a> Dec 31 1986 to Nov 15 2023
%<a href='https://www.macrotrends.net/2595/dax-30-index-germany-historical-chart-data'>DAX 30 Index - 27 Year Historical Chart</a> Nov 26 1990 to Nov 14 2023
%<a href='https://www.macrotrends.net/2596/cac-40-index-france-historical-chart-data'>CAC 40 Index - 27 Year Historical Chart</a> March 1 1990 to Nov 15 2023
%<a href='https://www.macrotrends.net/1320/nasdaq-historical-chart'>NASDAQ Composite - 45 Year Historical Chart</a>

\section{Results}\label{s:theory}
The goal is to understand the effect of $u$ and $v$ on 
\begin{equation*}
\max_LR_{n,r_1}^L-R_{n,r_0}^1,
\end{equation*}
which denotes the maximum difference in log-return between a daily leveraged ETF and its unlevered counterpart. Denote the $L$ that produces this maximum with $L^*$. In the following, an estimate of $L^*$ is produced that is a function of $u$ and $v$. 

First some intuition is developed. Recall that the Maclaurin series representation for $\log(1+x)$ is given by
\begin{equation*}
x-\frac{x^2}{2}+\frac{x^3}{3}-\frac{x^4}{4}+\ .\ .\ .\ 
\end{equation*}
When $x$ is sufficiently close to 0,
\begin{equation}\label{eq:logapprox}
\log(1+x)\approx x-\frac{x^2}{2}.
\end{equation}

Next, a function is introduced to more concisely describe the effect of fees on the difference in return between a leveraged ETF and its unlevered counterpart. Let $f:[0,1]^2\to\mathbb{R}$ be such that
\begin{equation*}
f(x,y):=\log\frac{1-y/252}{1-x/252}.
\end{equation*}
Note that $f(x,y)=-f(y,x)$ and $f(x,y)\approx (x-y)/252$. In particular, for $(x,y)\in[0,.05]^2$ there is
\begin{equation*}
\big\vert f(x,y)-\frac{x-y}{252}\big\vert\leq 2\cdot 10^{-8}.
\end{equation*} 

\begin{figure}
  \includegraphics[width=\linewidth]{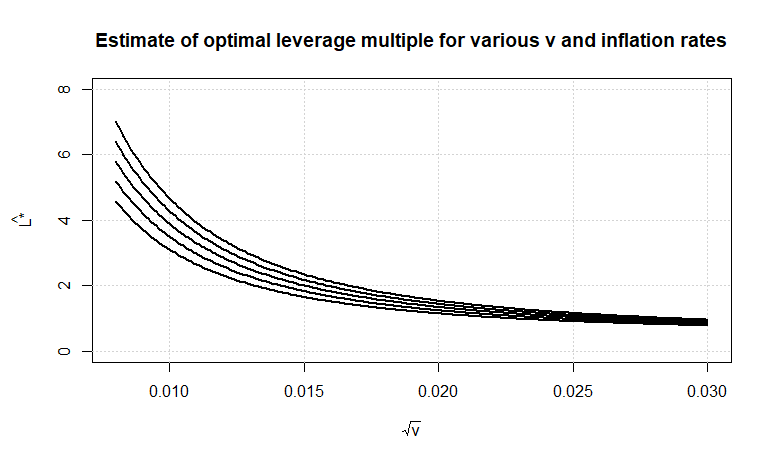}
  \caption{Illustrates $\hat{L^*}:=u/v+1/2$, which maximizes $g$. The $u$ are computed using table \ref{t:posrr}. For example, if the annual inflation rate is 2\%, then $u=.0856/252$. Each curve corresponds to an inflation rate in table \ref{t:posrr}. More specifically, the curves (from top to bottom) correspond to annual inflation rates of 4\%, 3\%, 2\%, 1\% and 0\%.}
  \label{fig:Ov}
\end{figure}

If the compounded error is sufficiently small, then \eqref{eq:RnrL} and \eqref{eq:logapprox} imply
\begin{equation*}
\begin{split}
R_{n,r_1}^L-R_{n,r_0}^1&=\Big[\sum_{i=1}^n\log(1+LX_i)-\log(1+X_i)\Big]+n\cdot f(r_0,r_1)\\
&\approx\Big[\sum_{i=1}^nLX_i-\frac{(LX_i)^2}{2}-\Big(X_i-\frac{X_i^2}{2}\Big)\Big]+n\cdot f(r_0,r_1)\\
&=(L-1)\Big[\sum_{i=1}^nX_i-\frac{L+1}{2}\cdot X_i^2\Big]+n\cdot f(r_0,r_1)\\
&\approx(L-1)\Big[\sum_{i=1}^n\log(1+X_i)-\frac{L}{2}\cdot X_i^2\Big]+n\cdot f(r_0,r_1)\\
&=(L-1)\Big[R_{n,0}^1-\sum_{i=1}^n\frac{L}{2}\cdot X_i^2\Big]+n\cdot f(r_0,r_1).
\end{split}
\end{equation*}
It follows that 
\begin{equation}\label{eq:RgLf}
R_{n,r_1}^L-R_{n,r_0}^1\approx n[g(L)+f(r_0,r_1)],
\end{equation}
where $g:\mathbb{R}\to\mathbb{R}$ is such that
\begin{equation*}
g(L):=(L-1)(u-Lv/2).
\end{equation*}
So roughly speaking,
\begin{equation}\label{eq:iffapprox}
R_{n,r_1}^L\leq R_{n,r_0}^1 \underset{\sim}{\iff} g(L)\leq f(r_1,r_0).
\end{equation}
Observe that
\begin{equation*}
\frac{\partial}{\partial L}g(L)=u+(1-2L)v/2.
\end{equation*}
Assuming $v>0$, it follows that the global maximum of $g$ is 
\begin{equation}\label{eq:h}
\frac{v}{2}\Big(\frac{u}{v}-\frac{1}{2}\Big)^2.
\end{equation}
and it occurs at $\hat{L^*}:=u/v+1/2$. Figure \ref{fig:Ov} shows how this estimate of the optimal leverage multiple is affected by volatility and inflation.

Let $h:(0,\infty)\to[0,\infty)$ be such that $h(v)$ is given by \eqref{eq:h}. Then
\begin{equation*}
\frac{\partial}{\partial v}h(v)=-\frac{1}{2}\Big(\frac{u^2}{v^2}-\frac{1}{4}\Big).
\end{equation*}
It follows that the global minimum of $h$ is 
\begin{equation*}
\begin{cases}
0& u>0\\
\vert u\vert & u<0
\end{cases},
\end{equation*}
and it occurs at $v=2\vert u\vert$. Moreover, $h$ is decreasing for $v<2\vert u\vert$ and increasing for $v>2\vert u\vert$. These results about $h$ can be interpreted as follows, assuming the underlying approximations hold. When $u>0$, there is a $v$ such that no daily leveraged index outperforms its unlevered counterpart. When $u<0$, for each $v$ there is at least one daily leveraged index outperforming the unlevered benchmark index.

\begin{figure}
  \includegraphics[width=\linewidth]{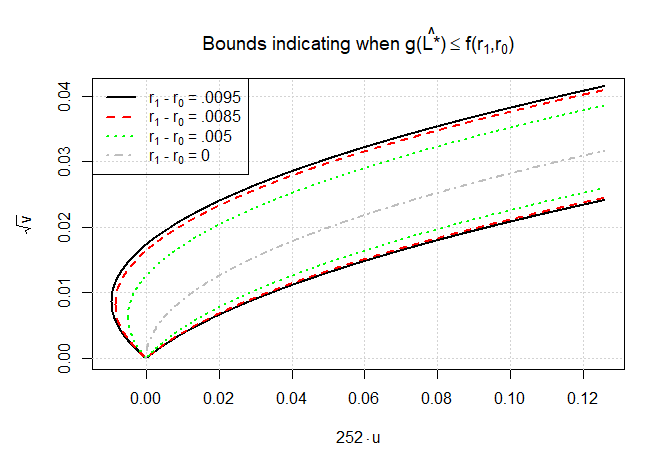}
  \caption{Recall \eqref{eq:vbounds}. $\sqrt{v^+}$ and $\sqrt{v^-}$ are the top and bottom curves, respectively. Only one curve is shown for $r_1-r_0=0$ because $v^+=v^-$. Note that $f(r_1,r_0)\approx (r_1-r_0)/252$, so different combinations of $r_1$ and $r_0$ can be identified using their difference instead. The average annualized log-return is given by $252\cdot u$.}
  \label{fig:vpvm}
\end{figure}

Next, the goal is to determine $v$ such that $g(\hat{L^*})\leq f(r_1,r_0)$. Performing some algebra and then applying the quadratic formula,
\begin{equation*}
\begin{split}
g(\hat{L^*})=f(r_1,r_0)&\iff \sqrt{\frac{v}{2}}\cdot\Big\vert\frac{u}{v}-\frac{1}{2}\Big\vert=\sqrt{f(r_1,r_0)}\\
&\iff\Big\vert u-\frac{v}{2}\Big\vert=\sqrt{2v\cdot f(r_1,r_0)}\\
&\iff \mp\frac{1}{2}\cdot v-\sqrt{2f(r_1,r_0)}\cdot \sqrt{v}\pm u=0\\
&\iff v=2\big(\sqrt{f(r_1,r_0)}\pm\sqrt{f(r_1,r_0)+u}\ \big)^2.%\\&\iff v=2\big(2f(r_1,r_0)+u\pm2\sqrt{f(r_1,r_0)(f(r_1,r_0)+u)}\ \big).
\end{split}
\end{equation*}
%It follows that $h(v)\leq f(r_1,r_0)$ if and only if
%\begin{equation*}
%-\sqrt{f(r_1,r_0)(f(r_1,r_0)+u)}\leq\frac{v}{4}-\frac{u}{2}-f(r_1,r_0)\leq\sqrt{f(r_1,r_0)(f(r_1,r_0)+u)}.
%\end{equation*}
Let $v^-$ and $v^+$ be such that
\begin{equation*}
v^{\pm}=2\big(\sqrt{f(r_1,r_0)}\pm\sqrt{f(r_1,r_0)+u}\ \big)^2.
\end{equation*}
Then 
\begin{equation}\label{eq:vbounds}
g(\hat{L^*})\leq f(r_1,r_0)\iff v\in[v^-,v^+].%v^-\leq v\leq v^+.
\end{equation}
Figure \ref{fig:vpvm} illustrates $v^-$ and $v^+$ for various $u$ and $r_1-r_0$. Note that $v^-$ and $v^+$ are undefined when $u<f(r_0,r_1)$, in which case $g(\hat{L^*})>f(r_1,r_0)$ for all $v>0$. 

Recalling \eqref{eq:RgLf}, there will (ideally) be 
\begin{equation}\label{eq:ngRR}
n\cdot g(\hat{L^*})\approx R_{n,0}^{L^*}-R_{n,0}^1,
\end{equation}
in which case \eqref{eq:vbounds} implies that
\begin{equation}\label{eq:ngRR2}
R_{n,r_1}^{L^*}\leq R_{n,r_0}^1 \underset{\sim}{\iff} v\in[v^-,v^+].
\end{equation}
So, ideally, the potential for a daily leveraged ETF to outperform its unlevered counterpart is determined by whether $v\in[v^-,v^+]$.

\subsection{Applications involving the S\&P 500 and optimal leverage multiples}\label{s:olm}
\begin{figure}
  \includegraphics[width=\linewidth]{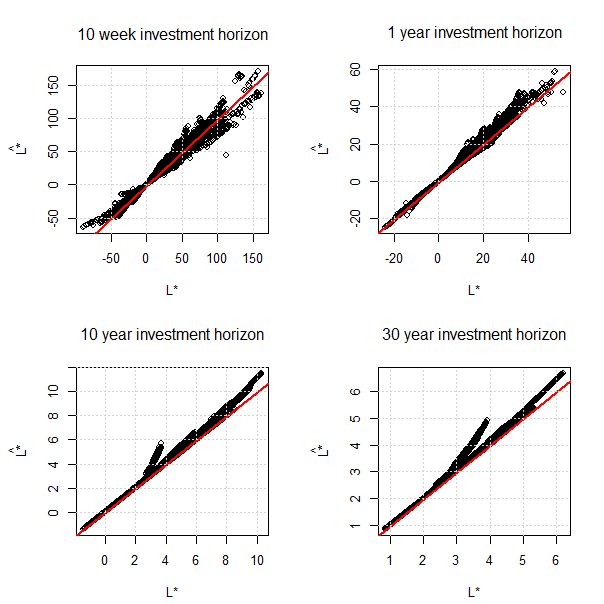}
  \caption{Using the S\&P 500 data, compares the optimal leverage multiple, $L^*$ (found numerically), with its estimate, $\hat{L^*}$. Note that $n=10\cdot 5,\ 252,\ 10\cdot 252,\ 30\cdot 252$ for 10 week, 1 year, 10 year and 30 year investment horizons. Each point represents a different starting date. The data spans December 29, 1927 to September 29, 2023. For example, the 10 week investment horizon has starting dates from December 29, 1927 to 10 weeks before September 29, 2023.}
  \label{fig:llhat}
\end{figure}

\begin{figure}
  \includegraphics[width=\linewidth]{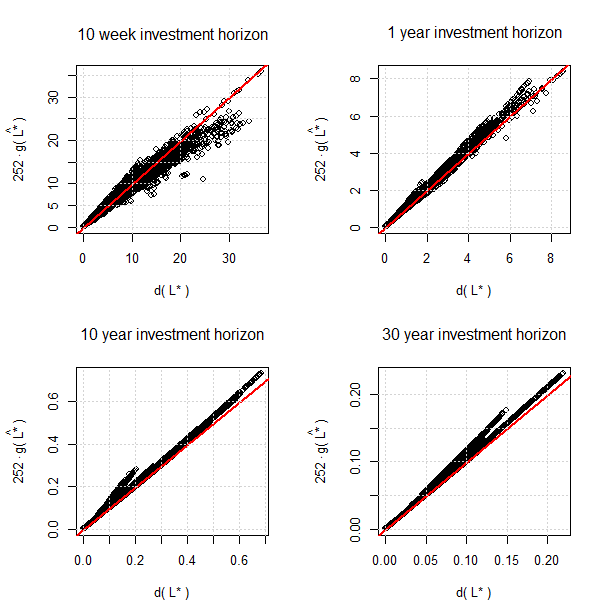}
  \caption{Using the S\&P 500 data, compares the maximum annualized return difference, $d(L^*)$ (found numerically), with its estimate, $252\cdot g(\hat{L^*})$. Note that $n=10\cdot 5,\ 252,\ 10\cdot 252,\ 30\cdot 252$ for 10 week, 1 year, 10 year and 30 year investment horizons. Each point represents a different starting date. The data spans December 29, 1927 to September 29, 2023. For example, the 10 week investment horizon has starting dates from December 29, 1927 to 10 weeks before September 29, 2023.}
  \label{fig:dg}
\end{figure}

Here, the estimate, $\hat{L^*}$, of the optimal leverage multiple, $L^*$, is applied to the S\&P 500 data. The goal is to examine the behavior of $\hat{L^*}$ versus $L^*$ and check the validity of
\begin{equation}\label{eq:goal2}
R_{n,r_1}^{L^*}\leq R_{n,r_0}^1 \underset{\sim}{\iff} g(\hat{L^*})\leq f(r_1,r_0)
\end{equation}
for relevant values of $r_0$ and $r_1$. The popular daily leveraged ETFs have fees no greater than 1\%, so relevant $r_0$ and $r_1$ are between $0$ and $.01$. Figures \ref{fig:llhat}, \ref{fig:dg}, \ref{fig:dgminus} and \ref{fig:dgminus2} illustrate the extent to which this goal is accomplished. Overall, $\hat{L^*}$ and $252\cdot g(\hat{L^*})$ provide decent estimations of $L^*$ and $d(L^*)$, respectively.

First consider figure \ref{fig:llhat}. For shorter time horizons like 10 weeks, the optimal leverage multiple, $L^*$, ranges from $-88$ to $162$. Leverage multiples outside the interval $[-5,5]$ are hardly practical, but it is still interesting to see what values $L^*$ can take on. The range of $L^*$ generally decreases as the time horizon increases: $[-23,56]$ for 1 year, $[-1.4,10.3]$ for 10 years and $[.84,6.22]$ for 30 years. This stabilization of $L^*$ is largely a result of the stabilization in $u$ and $v$ for longer time horizons. In words, the mean daily log-return and volatility in daily returns of the S\&P 500 become diluted as the time horizon increases, leading to less extreme optimal leverage multiples. 

Next consider figure \ref{fig:dg}. Regardless of time horizon, $252\cdot g(\hat{L^*})$ is highly correlated with $d(L^*)$, providing, at minimum, a ballpark estimate of $d(L^*)$. The relative spread between $252\cdot g(\hat{L^*})$ and $d(L^*)$ generally decreases as the time horizon increases. Observe from figure \ref{fig:llhat} that $\hat{L^*}$ approximates $L^*$ fairly well, tending to be an overestimate for time horizons of 1 year and beyond. As a result, $252\cdot g(\hat{L^*})$ also tends to overestimate $d(L^*)$ for time horizons of 1 year and beyond. So using $252\cdot g(\hat{L^*})$ can result in an overly optimistic idea of the optimal return difference $d(L^*)$. 

Recall from section \ref{s:theory} that the difference between $252\cdot f(r_1,r_0)$ and $r_1-r_0$ is negligible for $r_0,r_1\in[0,.01]$. So validation of \eqref{eq:goal2} reduces to checking
\begin{equation}\label{eq:goal3}
d(L^*)\leq r_1-r_0 \underset{\sim}{\iff} 252\cdot g(\hat{L^*})\leq r_1-r_0
\end{equation}
for $r_1-r_0\in[-.01,.01]$. Since $d(L^*)$ and $g(\hat{L^*})$ are nonnegative, \eqref{eq:goal3} holds for $r_1<r_0$. From here, the goal is to validate \eqref{eq:goal3} for $r_1-r_0\in[0,.01]$. Consider figures \ref{fig:dgminus} and \ref{fig:dgminus2}. Excluding three outliers, $\vert d(L^*)-252\cdot g(\hat{L^*})\vert\leq.002$ for 10 week and 1 year periods, provided $d(L^*)\leq.01$ or $252\cdot g(\hat{L^*})\leq.01$. For 10 year and 30 year periods, $\vert d(L^*)-252\cdot g(\hat{L^*})\vert\leq.0006$ when $d(L^*)\leq.01$ or $252\cdot g(\hat{L^*})\leq.01$. So in the instances where \eqref{eq:goal3} does not hold and $r_1-r_0\in[0,.01]$, the estimated optimal annualized return difference, $252\cdot g(\hat{L^*})-(r_1-r_0)$, differs from the actual optimal annualized return difference, $d(L^*)-(r_1-r_0)$, by at most .002 for shorter time horizons, and .0006 for longer time horizons. 

In conclusion, it is safe to rely on \eqref{eq:goal2}, because when it does not hold, the increase or decrease in return resulting from using $L=L^*$ instead of $L=1$ is hardly worthwhile. So if the decision to use $L=1$ or $L=L^*$ is based on whether $g(\hat{L^*})\leq f(r_1,r_0)$, then in the instances where the wrong choice is indicated, the cost is small.

\subsection{Applications involving the S\&P 500 and arbitrary leverage multiples}
Here, the goal is to show how well \eqref{eq:RgLf} holds, which is equivalent to checking
\begin{equation}\label{eq:d252gl}
d(L)\approx 252\cdot g(L).
\end{equation}
Figures \ref{fig:10wkArb} and \ref{fig:10yrArb} illustrate \eqref{eq:d252gl} for the common leverage multiples $L=-3,\ -2,\ -1,\ .5,\ 2$ and $3$. The difference between $d(L)$ and $252\cdot g(L)$ is generally smaller for longer time horizons, like 10 years instead of 10 weeks. The difference can be quite large for shorter time horizons like 10 weeks if $\vert L\vert\geq2$; but relative to $d(L)$, there are fewer instances where that difference is extreme. So \eqref{eq:d252gl} holds up well under long time horizons, but starts to break down under short time horizons with high magnitude leverage multiples. In other words, $u$ and $v$ can be used to achieve a decent prediction of the return of a daily leveraged ETF, especially for longer time horizons and/or lower magnitude leverage multiples. Thus, an accurate prediction of $u$ and $v$ can yield a decent prediction of the return for a daily leveraged S\&P 500 ETF. 

\subsection{Effect of higher moments}\label{s:hm}
The goal here is to determine the effect of incorporating higher moments, which impact skewness and kurtosis, into the estimation of daily leveraged ETF log-returns. First observe that 
\begin{equation}\label{eq:logapprox2}
\log(1+x)\approx x-\frac{x^2}{2}+\frac{x^3}{3}-\frac{x^4}{4}.
\end{equation}
If the compounded error coming from higher moments (beyond skewness and kurtosis) is sufficiently small, then \eqref{eq:RnrL} and \eqref{eq:logapprox2} imply
{\allowdisplaybreaks
\begin{align*}
R_{n,r_1}^L-R_{n,r_0}^1&=\Big[\sum_{i=1}^n\log(1+LX_i)-\log(1+X_i)\Big]+n\cdot f(r_0,r_1)\\
&\approx\Big[\sum_{i=1}^nLX_i-\frac{(LX_i)^2}{2}+\frac{(LX_i)^3}{3}-\frac{(LX_i)^4}{4}\\
&\quad\quad-\Big(X_i-\frac{X_i^2}{2}+\frac{X_i^3}{3}-\frac{X_i^4}{4}\Big)\Big]+n\cdot f(r_0,r_1)\\
&=(L-1)\Big[\sum_{i=1}^nX_i-\frac{L+1}{2}\cdot X_i^2+\frac{L^2+L+1}{3}\cdot X_i^3\\
&\quad\quad\quad\quad\quad-\frac{L^3+L^2+L+1}{4}\cdot X_i^4\Big]+n\cdot f(r_0,r_1)\\
&\approx(L-1)\Big[\sum_{i=1}^n\log(1+X_i)-\frac{L}{2}\cdot X_i^2+\frac{L^2+L}{3}\cdot X_i^3\\
&\quad\quad\quad\quad\quad-\frac{L^3+L^2+L}{4}\cdot X_i^4\Big]+n\cdot f(r_0,r_1)\\
&=(L-1)\Big[R_{n,0}^1-L\Big(\sum_{i=1}^n\frac{1}{2}\cdot X_i^2-\frac{L+1}{3}\cdot X_i^3\\
&\quad\quad\quad\quad\quad\quad\quad\quad\quad\quad+\frac{L^2+L+1}{4}\cdot X_i^4\Big)\Big]+n\cdot f(r_0,r_1).
\end{align*}
}
Let 
\begin{equation*}
m_3=\frac{1}{n}\sum_{i=1}^nX_i^3\ ,\quad m_4=\frac{1}{n}\sum_{i=1}^nX_i^4.
\end{equation*}
It follows that 
\begin{equation}\label{eq:approxggtilde}
R_{n,r_1}^L-R_{n,r_0}^1\approx n[g(L)+\widetilde{g}(L)+f(r_0,r_1)],
\end{equation}
where
\begin{equation*}
\widetilde{g}(L)=\frac{L^3-L}{3}\cdot m_3-\frac{L^4-L}{4}\cdot m_4.
\end{equation*}

Validation of \eqref{eq:approxggtilde} reduces to checking  
\begin{equation}\label{eq:approxggtilde2}
d(L)\approx 252\cdot [g(L)+\widetilde{g}(L)].
\end{equation}
Figures \ref{fig:10wkArb34} and \ref{fig:10yrArb34} are the analogues of figures \ref{fig:10wkArb} and \ref{fig:10yrArb}, designed to illustrate the accuracy of \eqref{eq:approxggtilde2}. It is not hard to see that $252\cdot [g(L)+\widetilde{g}(L)]$ is generally closer to $d(L)$ than $252\cdot g(L)$. However, recall that $252\cdot g(L)$ was already quite close to $d(L)$ for longer time horizons like 10 years. The main benefit of using $252\cdot [g(L)+\widetilde{g}(L)]$ instead of $252\cdot g(L)$ has to do with shorter time horizons and high magnitude leverage multiples, where the difference between $252\cdot g(L)$ and $d(L)$ can be relatively large. 

Figures \ref{fig:lltilde} and \ref{fig:dg34} are the analogues of figures \ref{fig:llhat} and \ref{fig:dg}, designed to compare $\tilde{L^*}$ with $L^*$. In general, $\tilde{L^*}$ and $g(\tilde{L^*})+\widetilde{g}(\tilde{L^*})$ are improvements over $\hat{L^*}$ and $g(\hat{L^*})$, respectively. However, as detailed in section \ref{s:olm}, \eqref{eq:goal2} is already reliable. Thus, the benefit gained from instead using \eqref{eq:goal22} is minimal. Moreover, \eqref{eq:goal22} requires knowledge of $u$, $v$, $m_3$ and $m_4$, whereas \eqref{eq:goal22} requires knowledge of just $u$ and $v$. So the small improvement in accuracy offered by \eqref{eq:goal22} over \eqref{eq:goal2} is seemingly not worth the added difficulty associated with predicting $m_3$ and $m_4$.
\begin{equation}\label{eq:goal22}
R_{n,r_1}^{L^*}\leq R_{n,r_0}^1 \underset{\sim}{\iff} g(\tilde{L^*})+\widetilde{g}(\tilde{L^*})\leq f(r_1,r_0)
\end{equation}

To summarize, if one is able to reliably predict $m_3$ and $m_4$, then the use of $\tilde{L^*}$ and $\tilde{g}$ may be warranted. Their use does not seem worthwhile for the boolean determination of whether a daily leveraged ETF can outperform its unlevered counterpart, since $\hat{L^*}$ and $g$ are already quite successful in this capacity (see section \ref{s:olm}). However, in estimating a daily leveraged ETF's future return, $\tilde{g}$ can offer a significant improvement to accuracy, especially for short time horizons or high magnitude leverage multiples. 

\subsection{Prediction of estimator inputs for the S\&P 500}
Here, the predictability of $u$, $v$, $m_3$ and $m_4$ is discussed for the S\&P 500 Index. Looking at figure \ref{fig:u}, $252\cdot u$ fluctuates between positive and negative values for time horizons of 10 weeks, 1 year and 10 years. However, $252\cdot u$ stabilizes at the 30 year investment horizon, remaining above .05. For multi-decade investment horizons, inflation and the long-term positive trend shown in figure \ref{fig:sptrend} keeps $252\cdot u$ from dipping below .05. Thus, it is reasonable to expect $252\cdot u$ to lie between $.05$ and $.1$ for long investment horizons. For short horizons, a decent prediction of $u$ seems less attainable. One would have to predict short-term market trends, which is extremely difficult. 

Looking at figures \ref{fig:v} and \ref{fig:m4}, the range of $\sqrt{v}$ and $\sqrt[4]{m_4}$ decreases as the investment horizon increases. An upward trend in $\sqrt{v}$ is apparent for the 10 and 30 year investment horizons. Thus, it seems reasonable to predict that $\sqrt{v}$ will remain above $.01$ for long investment horizons, possibly exceeding .015, or even .02, if the upward trend continues. An upward trend in $\sqrt[4]{m_4}$ is also noticeable for the 10 year investment horizon; however, it is not so clear for the 30 year investment horizon. Thus it is difficult to expect much beyond having $\sqrt[4]{m_4}$ lie somewhere between .01 and .03 for 10 year horizons, or between .02 and .026 for 30 year horizons.

Looking at figures \ref{fig:m3} and \ref{fig:m4}, $\sqrt[3]{m_3}$ and $\sqrt[4]{m_4}$ can have surprising jumps regardless of investment horizon. These jumps are a result of rare extremes in the daily percentage changes of the S\&P 500. For example, there was a -20.5\% change on October 19 1987 (Black Monday). Thus, accurate prediction of $m_3$ and $m_4$ has a lot to do with anticipation of rare extremes in the daily percentage changes, which seems far-fetched. For long investment horizons, the range of $m_4$ appears to be more stable than that of $m_3$. However, according to the methods in section \ref{s:hm}, reliance on $m_4$ to estimate daily leveraged ETF returns necessarily involves $m_3$. Since $m_3$ is apparently difficult to predict, the estimation methodology of section \ref{s:hm} is not recommended.

In summary, rough prediction of $u$ and $v$ seems reasonable for holding periods spanning multiple decades. In particular, $252\cdot u$ can be bounded between $.05$ and $.1$, and $\sqrt{v}$ can be given a lower bound of .01. Moreover, it appears possible for $\sqrt{v}$ to climb above .015, and maybe even .02 eventually. The fluctuations of $u$ and $v$ associated with short investment horizons makes them difficult to predict in those settings. Prediction of $m_3$ is questionable regardless of investment horizon, making estimations of daily leveraged ETF returns that involve higher moments unreliable, since they likely use $m_3$.

\subsection{Application to other indexes}\label{s:otheridx}
Here, the goal is to check the general validity of approximating $d(L)$ with $252\cdot g(L)$. Results indicate how well that approximation applies to arbitrary indexes. In the following, two linear programs (LPs) are built in order to bound $d(L)$, provided incomplete information about the daily percentage changes of the underlying benchmark index. Then it is possible to compare $252\cdot g(L)$ with those bounds. 

To be clear, the incomplete information about the underlying benchmark index's daily percentage changes is:
\begin{itemize}
\item $u$ and $v$ are known,
\item $m_3\in[\underline{m_3},\overline{m_3}]$ and $m_4\in[\underline{m_4},\overline{m_4}]$, where $\underline{m_3},\ \overline{m_3},\ \underline{m_4}$ and $\overline{m_4}$ are known,
\item $X_i\in[\underline{z},\overline{z}]$ for $i=1,2,...,n$, where $\underline{z}$ and $\overline{z}$ are known.
\end{itemize}

Now, the LPs are built in order to bound $d(L)$. First choose $\underline{z},\overline{z}\in\mathbb{R}$ such that 
\begin{itemize}
\item $-1<\underline{z}<0<\overline{z}$,
\item $\log(1+Lx)$ exists for all $x\in[\underline{z},\overline{z}]$,
\item $X_i\in[\underline{z},\overline{z}]$ for each $i=1,2,...,n$.
\end{itemize}
In words, $\underline{z}$ and $\overline{z}$ represent lower and upper bounds on the daily percentage changes of the underlying benchmark index.

Next choose 
\begin{equation*}
\boldsymbol\delta=(\delta_1,\delta_2,\delta_3,\delta_4,\delta_5)^T\in(0,\infty)^5. 
\end{equation*}
According to section \ref{a:bounds} and algorithm \ref{alg:zi}, build $\mathbf{z}=(z_1,z_2,...,z_m)^T\in[\underline{z},\overline{z}]^m$. 

For $j=1,2,...,m$, let 
\begin{align*}
c_j&=252\cdot\log\frac{1+Lz_j}{1+z_j}, &a_{1j}&=1,& a_{2j}&=\log(1+z_j),\\
\quad a_{3j}&=z_j^2,& a_{4j}&=z_j^3,& a_{5j}&=z_j^4.
\end{align*}
Let $\mathbf{g},\mathbf{c}\in\mathbb{R}^m$, $\mathbf{A}=(a_{ij})$ and 
\begin{equation*}
\begin{split}
\underline{\mathbf{b}}^T&=(1,u,v,\underline{m_3},\underline{m_4}),\\
\overline{\mathbf{b}}^T&=(1,u,v,\overline{m_3},\overline{m_4}).
\end{split}
\end{equation*}
Note that $\underline{\mathbf{b}}$ and $\overline{\mathbf{b}}$ are extended real vectors, meaning some of their entries can be $\pm\infty$. The LPs of interest are
\begin{subequations}
\begin{align}
\min\ \mathbf{c}^T\mathbf{g}\text{ s.t. }\underline{\mathbf{b}}-\boldsymbol\epsilon\leq\mathbf{Ag}\leq\overline{\mathbf{b}}+\boldsymbol\epsilon,\ \mathbf{g}\geq\mathbf{0},&\tag{\underline{LP}}\label{eq:min}\\
\max\ \mathbf{c}^T\mathbf{g}\text{ s.t. }\underline{\mathbf{b}}-\boldsymbol\epsilon\leq\mathbf{Ag}\leq\overline{\mathbf{b}}+\boldsymbol\epsilon,\ \mathbf{g}\geq\mathbf{0},&\tag{$\overline{\text{LP}}$}\label{eq:max}
\end{align}
\end{subequations}
where $\boldsymbol\epsilon=(0,\delta_1,\delta_2,\delta_3,\delta_4)^T$. In case it is not clear, the inequalities in \eqref{eq:min} and \eqref{eq:max} must be satisfied element-wise. 

In words, \eqref{eq:min} and \eqref{eq:max} optimize 
\begin{equation*}
252\cdot\mathbb{E}[\log(1+LZ)-\log(1+Z)]
\end{equation*}
subject to constraints on $\mathbb{E}[\log(1+Z)]$, $\mathbb{E}[Z^2]$, $\mathbb{E}[Z^3]$ and $\mathbb{E}[Z^4]$, where $\mathbb{E}$ denotes the expectation, $Z$ is a discrete random variable taking on values in the set $\{z_1,z_2,...,z_m\}$ and $Z$ represents the daily percentage changes of the underlying benchmark index. Denote the optimal objective function value of a given LP, if it exists, as that LP's equation reference with a superscript $*$. For example, the optimal objective function value of \eqref{eq:min} would be denoted $\eqref{eq:min}^*$.

Observe that \eqref{eq:min} and \eqref{eq:max} are bounded because the $c_j$ and $g_j$ are bounded. So assuming feasibility, the optimal objective function values exist and $\eqref{eq:min}^*\leq\eqref{eq:max}^*$.

Now here is the \textbf{main result}: if $m_3\in[\underline{m_3},\overline{m_3}]$ and $m_4\in[\underline{m_4},\overline{m_4}]$, then provided feasibility,
\begin{equation}\label{eq:mr}
\eqref{eq:min}^*-252\cdot(\delta_1+\delta_5)\leq d(L)\leq \eqref{eq:max}^*+252\cdot(\delta_1+\delta_5).
\end{equation}
For an explanation of \eqref{eq:mr}, see section \ref{a:dLbounds}.
%for any $\delta_5>0$. %In applications, the choice of $\delta_5=10^{-5}/252$ was made arbitrarily, with the goal of keeping the bounds on $d(L)$ tight, but not blowing up $m$. %It is possible to replace $20^{-5}$ in \eqref{eq:mr} with another positive number closer to 0, but the benefit of doing so is likely negligible for practical purposes.

For applications, the choice of 
\begin{equation}\label{eq:deltachoice}
\boldsymbol\delta=(10^{-5}/252,\ 10^{-6},\ 10^{-8},\ 10^{-10},\ 10^{-5}/252)^T
\end{equation}
was made arbitrarily, with the goal of keeping $\boldsymbol\delta$ close to $\mathbf{0}$, but not blowing up $m$. In general, $m$ increases as $\boldsymbol\delta$ approaches $\mathbf{0}$. As shown previously, solutions (assuming they exist) to \eqref{eq:min} and \eqref{eq:max} bound $d(L)$. So having $\boldsymbol\delta$ close to $\mathbf{0}$ creates tighter bounds. 

Additionally, applications use
\begin{equation}\label{eq:m3m4app}
\underline{m_3}=-.02^3,\ \overline{m_3}=.02^3,\ \underline{m_4}=0,\ \overline{m_4}=.04^4.
\end{equation}
Note that $\boldsymbol\delta,\underline{m_3},\overline{m_3},\underline{m_4},\overline{m_4}$ were chosen arbitrarily to achieve practically worthwhile results. In particular, $\underline{m_3},\overline{m_3},\underline{m_4},\overline{m_4}$ align with historical norms of $m_3$ and $m_4$ for large-cap market indexes over longer time spans like 10+ years (see figures \ref{fig:m3}, \ref{fig:m4} and \ref{fig:fghs2}). 

Table \ref{t:m} gives an idea of how large $m$ is for some reasonable $\underline{z}$ and $\overline{z}$. Daily percentage changes outside of $\pm25\%$ are unheard of for many large-cap market indexes like the S\&P 500 and DJIA. Smaller markets have more potential to deviate from this norm, but only on extremely rare occasions. Most importantly, $m$ is not too large, practically speaking, to use in computation of \eqref{eq:min} and \eqref{eq:max}, where $m$ serves as the number of columns. 
\begin{table}
\begin{center}
\caption{Gives the $m$ produced from algorithm \ref{alg:zi} when $\underline{z}=-\overline{z}$ and $\boldsymbol\delta$ is as in \eqref{eq:deltachoice}.}\label{t:m}
\begin{tabular}{ c|cccccc } 
 \bottomrule
 $L$ & -3 & -2 & -1 & .5 & 2 & 3 \\ \hline
 $m$ ($\overline{z}=.25$)& 8845 & 8698 & 8612 & 8612 & 8698 & 8844 \\ 
 $m$ ($\overline{z}=.35$) & 10278 & 10132 & 10046 & 10046 & 10132 & NA \\ 
 \toprule
\end{tabular}
\end{center}
\end{table}

Tables \ref{t:n3}, \ref{t:n2}, \ref{t:n1}, \ref{t:p5}, \ref{t:2} and \ref{t:3} show how well $252\cdot g(L)$ approximates $d(L)$, by use of \eqref{eq:mr}. For some of the most practical leverage multiples ($L=-3,-2,-1,.5,2,3$), $d(L)$ aligns with $252\cdot g(L)$, provided $\overline{z}=-\underline{z}=.25$, $\boldsymbol\delta$ is as in \eqref{eq:deltachoice}, $m_3\in[\underline{m_3},\overline{m_3}]$, $m_4\in[\underline{m_4},\overline{m_4}]$ and $\underline{m_3},\overline{m_3},\underline{m_4},\overline{m_4}$ are as in \eqref{eq:m3m4app}. Here are some important observations pertinent to those leverage multiples.
\begin{itemize}
\item In general, $252\cdot g(L)$ is between the lower and upper bounds of $d(L)$, but tends to lie closer to the upper bound.
\item For relevant values of $u$ and $v$, the (absolute) difference between $252\cdot g(L)$ and $d(L)$ is at most .053 for $L=-3,3$, .009 for $L=-2,2$ and .001 for $L=-1,.5$. So large magnitude leverage multiples create more room for error between $252\cdot g(L)$ and $d(L)$.
\item The difference between $252\cdot g(L)$ and $d(L)$ depends mostly on $L$. So for fixed $L$, the difference hardly changes across relevant values of $u$ and $v$.
\end{itemize}

Overall, $252\cdot g(L)$ approximates $d(L)$ well in practical long-term investment situations (10+ years), where $m_3$ and $m_4$ are less extreme. So if one is debating what leverage multiple in $[-3,3]$ to employ over a long time horizon, decisions can be based on using $252\cdot g(L)$ as a proxy for $d(L)$. In other words, predictions of $u$ and $v$ are sufficient to produce an accurate prediction of $d(L)$, allowing the choice of $L$ to be based on those predictions of $u$ and $v$. For example, the tables show that
\begin{equation}\label{eq:ELv}
\text{Estimate}-\text{Lower Bound}=252\cdot g(L)-[\eqref{eq:min}^*-252\cdot(\delta_1+\delta_5)]
\end{equation}
depends almost entirely on just $L$ and $v$. So \eqref{eq:ELv} can be approximated very well as a function of $L$ and $v$, call it $E(L,v)$. As a result, a maxmin (robust) optimization of $d(L)$ (over $L$) can be carried out by simply maximizing $252\cdot g(L)-E(L,v)$ over $L$. 

Most importantly, the observation that $252\cdot g(L)$ lies between the bounds of $d(L)$, yet closer to the upper bound, makes it reasonable treat $252\cdot g(\hat{L^*})$ as $d(L^*)$. So if predictions of $u$ and $v$ indicate $252\cdot g(\hat{L^*})\leq0$, it is reasonable to have $d(L^*)\leq0$ as well. 

Out of curiosity, the author also examined the analogues of tables \ref{t:n3}, \ref{t:n2}, \ref{t:n1}, \ref{t:p5}, \ref{t:2} and \ref{t:3} using 
\begin{equation*}
\underline{m_3}=-\infty,\ \overline{m_3}=\infty,\ \underline{m_4}=0,\ \overline{m_4}=\infty.
\end{equation*}
The spread between the upper and lower bounds for $d(L)$ increased dramatically. So $252\cdot g(L)$ cannot be reliably used as a proxy for $d(L)$ when no information about $m_3$ and $m_4$ is assumed. However, the difference between the bounds and $252\cdot g(L)$ still appeared to be very well approximated as functions of $L$ and $v$. 

\section{Conclusion}\label{s:conclusion}
If volatility in daily returns continues to increase, eventually passing the necessary threshold, then daily leveraged S\&P 500 ETFs will become obsolete for long-term investment. If volatility remains below this threshold, there will be opportunity for leveraged S\&P 500 ETFs to beat standard S\&P 500 ETFs. However, it may be difficult to take advantage of these opportunities if future volatility becomes difficult to predict. %especially if no noticeable correlation exists between expected daily log-return and daily volatility.

The focus here was on the S\&P 500 because of its popularity and historic stability. Other indexes may also produce decent results. However, any meaningful application will need a sufficiently accurate prediction of the mean daily log-return for the period in question. Examples include an upper or lower bound on the mean that is anticipated to hold with high confidence. If such a prediction is not obtainable, then it may be wise to avoid any leveraged ETF based on that index. As mentioned in section \ref{s:otheridx}, large-cap market indexes are more likely to exhibit long-term stability. Some examples include the CAC 40 (France), DAX 30 (Germany) and HSI (Hong Kong).

A potential extension to this work involves considering shape constraints in the optimization of $\mathbb{E}[\log(1+LZ)]$ subject to constraints on $\mathbb{E}[\log(1+Z)]$, $\mathbb{E}[Z^2]$, $\mathbb{E}[Z^3]$ and $\mathbb{E}[Z^4]$ ($Z$ represents daily percentage changes of the underlying index). For example, one could assume that $Z$ is a continuous random variable with unimodal distribution. The goal would be to compare results with and without various shape constraints. Expectation optimization subject to moment and shape constraints is considered in \cite{chen2021discrete}. It may also be interesting to compare those results with the alternative formulation: optimize $\mathbb{E}[\log(1+L(\exp \hat{Z}-1))]$ subject to constraints on $\mathbb{E}[\hat{Z}]$, $\mathbb{E}[\hat{Z}^2]$, $\mathbb{E}[\hat{Z}^3]$ and $\mathbb{E}[\hat{Z}^4]$. In this formulation, $\hat{Z}$ represents the daily log returns of the underlying index.

As identified in \cite{washer2016increasing}, there is an interesting phenomenon occurring where volatility in daily returns increases, but volatility in returns over a longer period, say monthly, does not. This is not how returns should behave if daily returns are generally independent and identically distributed. Future research could investigate what sort of stochastic processes can exhibit increased short-term volatility while maintaining long-term volatility.

\begin{appendix}
\section{Computation of $m$ and $z_1,z_2,...,z_m$}\label{a:bounds}
Let $x\in\mathbb{R}$ and $\delta>0$. Suppose $\phi:[x,x+\delta]\to\mathbb{R}$ is differentiable and either strictly concave or strictly convex. Denote the derivative of $\phi$ as $\phi'$.%Further suppose that the derivative of $\phi$, denoted $\phi'$, is strictly monotone. 

Let $\psi_{x,\delta}:[0,\delta]\to\mathbb{R}$ be such that 
\begin{equation*}
\psi_{x,\delta}(t)=\phi(x)+\frac{t}{\delta}\cdot[\phi(x+\delta)-\phi(x)]-\phi(x+t).
\end{equation*}
It can be shown with basic calculus that $\vert\psi_{x,\delta}(t)\vert$ is maximized at $t=t^*$, where 
\begin{equation*}
t^*=(\phi')^{-1}\Big(\frac{\phi(x+\delta)-\phi(x)}{\delta}\Big)-x,
\end{equation*}
and $(\phi')^{-1}$ denotes the inverse of $\phi'$. Thus, the maximum (absolute) difference between $\phi(x+t)$ and 
\begin{equation*}
\Big(1-\frac{t}{\delta}\Big)\cdot\phi(x)+\frac{t}{\delta}\cdot\phi(x+\delta)
\end{equation*}
is $\vert\psi_{x,\delta}(t^*)\vert$. 

Observe that 
\begin{equation*}
\frac{\partial }{\partial \delta}\psi_{x,\delta}(t)=\frac{t}{\delta}\cdot\Big(\phi'(x+\delta)-\frac{\phi(x+\delta)-\phi(x)}{\delta}\Big).
\end{equation*}
So for fixed $x$ and $t$, $\vert\psi_{x,\delta}(t)\vert$ decreases when $\delta$ decreases, provided $t\leq\delta$ and the supposed properties of $\phi$ and $\phi'$ hold for the $\delta$ under consideration. This means that if $\vert\psi_{x,\delta}(t^*)\vert\leq C$ for some constant $C$, then $\vert\psi_{x,\delta_0}(t^*)\vert\leq C$ for all $\delta_0\in(0,\delta]$.

Algorithm \ref{alg:zi} uses the previous results of this section to generate a finite subset, $\{z_1,z_2,...,z_m\}$, of the interval $[\underline{z},\overline{z}]$ such that the linear interpolation connecting a given function's values on $\{z_1,z_2,...,z_m\}$ are sufficiently close to the actual function values on $[\underline{z},\overline{z}]$. So if the given function is $\phi_0:[\underline{z},\overline{z}]\to\mathbb{R}$, then the linear interpolation is $\hat{\phi}_0:[\underline{z},\overline{z}]\to\mathbb{R}$, where 
\begin{equation*}
\hat{\phi}_0(x)=\Big(1-\frac{t}{z_{j+1}-z_j}\Big)\cdot\hat{\phi}_0(z_j)+\frac{t}{z_{j+1}-z_j}\cdot\hat{\phi}_0(z_{j+1}),
\end{equation*}
with $t=x-z_j$ and $z_j$ the largest element of $\{z_1,z_2,...,z_m\}$ satisfying $t>0$. Algorithm \ref{alg:zi} determines $\{z_1,z_2,...,z_m\}$ such that the distance between $\phi_0(x)$ and $\hat{\phi}_0(x)$ does not exceed a given amount for all $x\in[\underline{z},\overline{z}]$. Note that algorithm \ref{alg:zi} uses
\begin{equation}\label{eq:phi0list}
\phi_0(x)=x^2,\ x^3,\ x^4,\ \log(1+x),\ \log(1+Lx),
\end{equation}
and relies on the fact that on each side of $x=0$, $\phi_0(x)$ is differentiable and either strictly concave or strictly convex. 

Note that choice of $\boldsymbol\delta$ in algorithm \ref{alg:zi} is arbitrary, and the algorithm will work with any $\boldsymbol\delta$ having all-positive entries. Be advised, as the entries of $\boldsymbol\delta$ get closer to 0, $m$ will increase, possibly to a level that is not practical. Proof that the algorithm will finish for any $\boldsymbol\delta>\mathbf{0}$ lies in the fact that the $\phi_0$ in \eqref{eq:phi0list} have bounded continuous second derivatives on $[\underline{z},\overline{z}]$, and their linear interpolations, $\hat{\phi}_0$, can be made arbitrarily close to $\phi_0$ by having a sufficiently fine mesh $\{z_1,z_2,...,z_m\}$ (see section 6.1 of \cite{kincaid2009numerical}).%can be proven by showing that for bounded functions on bounded interval, linear interpolation gets arbitrarily close to its parent function as mesh becomes finer and finer

\section{Explanation of LP bounds for $d(L)$}\label{a:dLbounds}
Explanation of \eqref{eq:mr} relies on section \ref{a:bounds} and algorithm \ref{alg:zi}. It can be shown that there is a feasible solution, $\hat{\mathbf{g}}$, to \eqref{eq:min} and \eqref{eq:max} such that 
\begin{equation}\label{eq:ctghatbound}
\vert\mathbf{c}^T\hat{\mathbf{g}}-d(L)\vert\leq 252\cdot(\delta_1+\delta_5).
\end{equation}
Specifically, $\hat{\mathbf{g}}=(\hat{g}_1,\hat{g}_2,...,\hat{g}_m)^T$ is given by
\begin{equation*}
\hat{g}_j=\frac{1}{n}\cdot\sum_{i=1}^n\mathcal{X}_{(z_{j-1},z_j)}(X_i)\cdot\frac{X_i-z_{j-1}}{z_j-z_{j-1}}+\mathcal{X}_{[z_{j},z_{j+1})}(X_i)\cdot\Big(1-\frac{X_i-z_j}{z_{j+1}-z_{j}}\Big),
\end{equation*}
where $z_0=z_1$, $z_{m+1}=z_m$ and $\mathcal{X}_A(x)$ denotes the indicator function for $A\subset\mathbb{R}$. 

To understand where $\hat{\mathbf{g}}$ comes from, first let $G:[\underline{z},\overline{z}]\to\mathbb{R}^6$ be such that
\begin{equation*}
G(x)=\Big(252\cdot\log\frac{1+Lx}{1+x},\ 1,\ \log(1+x),\ x^2,\ x^3,\ x^4\Big)^T.
\end{equation*}
Then 
\begin{equation}\label{eq:Gsum}
\frac{1}{n}\sum_{i=1}^n G(X_i)=(d(L),\ 1,\ u,\ v,\ m_3,\ m_4)^T
\end{equation}
is approximated via linear interpolation with $\frac{1}{n}\sum_{i=1}^n\hat{G}(X_i)$, where 
\begin{equation*}
\begin{split}
\hat{G}(X_i)&=\Big(1-\frac{X_i-z_{j_i}}{z_{j_i+1}-z_{j_i}}\Big)\cdot G(z_{j_i})+\frac{X_i-z_{j_i}}{z_{j_i+1}-z_{j_i}}\cdot G(z_{j_i+1})\\
j_i&=\min\{j\in\{1,2,...,m-1\}:\ X_i\in[z_j,z_{j+1}]\}.
\end{split}
\end{equation*}
Moreover, 
\begin{equation}\label{eq:ghatmatrix}
\frac{1}{n}\sum_{i=1}^n\hat{G}(X_i)=\begin{pmatrix}
\mathbf{c}^T\\
\mathbf{A}
\end{pmatrix}\cdot\hat{\mathbf{g}},
\end{equation}
and in algorithm \ref{alg:zi}, the $z_1,z_2,...,z_m$ were selected such that
\begin{equation}\label{eq:GGhat}
\begin{split}
\Big\vert\frac{1}{n}\sum_{i=1}^nG(X_i)-\frac{1}{n}\sum_{i=1}^n\hat{G}(X_i)\Big\vert\leq\ &\frac{1}{n}\cdot\sum_{i=1}^n\vert G(X_i)-\hat{G}(X_i)\vert\\
\leq\ &\frac{1}{n}\cdot\sum_{i=1}^n(252\cdot(\delta_1+\delta_5),\ 0,\ \delta_1,\ \delta_2,\ \delta_3,\ \delta_4)^T\\
=\ &(252\cdot(\delta_1+\delta_5),\ 0,\ \delta_1,\ \delta_2,\ \delta_3,\ \delta_4)^T
\end{split}
\end{equation}
Note that for $\mathbf{v}\in\mathbb{R}^6$, $\vert\mathbf{v}\vert=(\vert v_1\vert,\vert v_2\vert,...,\vert v_6\vert)^T$. Observe that \eqref{eq:ctghatbound} now follows from \eqref{eq:Gsum}, \eqref{eq:ghatmatrix} and \eqref{eq:GGhat}.

\end{appendix}

%%===========================================================================================%%
%% If you are submitting to one of the Nature Portfolio journals, using the eJP submission   %%
%% system, please include the references within the manuscript file itself. You may do this  %%
%% by copying the reference list from your .bbl file, paste it into the main manuscript .tex %%
%% file, and delete the associated \verb+\bibliography+ commands.                            %%
%%===========================================================================================%%
\clearpage
\bibliography{sn-bibliography}% common bib file
%% if required, the content of .bbl file can be included here once bbl is generated
%%\input sn-article.bbl

%% Default %%
%%\input sn-sample-bib.tex%

\clearpage

\begin{figure}
  \includegraphics[width=\linewidth]{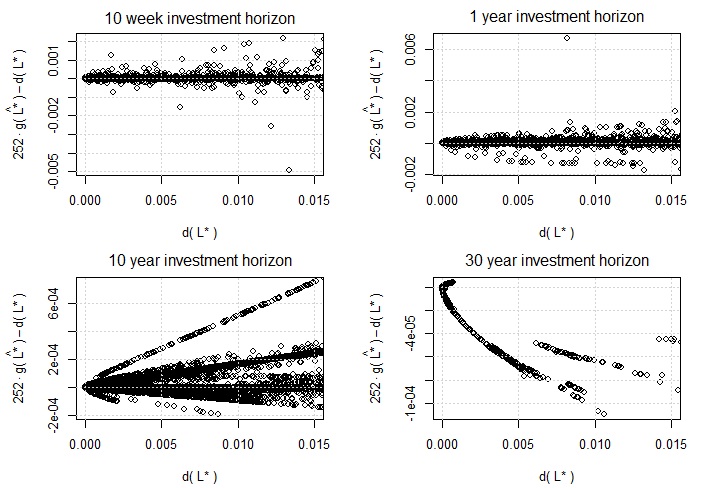}
  \caption{Like figure \ref{fig:dg}, but constricted to $d(L^*)\leq .015$.}
  \label{fig:dgminus}
\end{figure}

\begin{figure}
  \includegraphics[width=\linewidth]{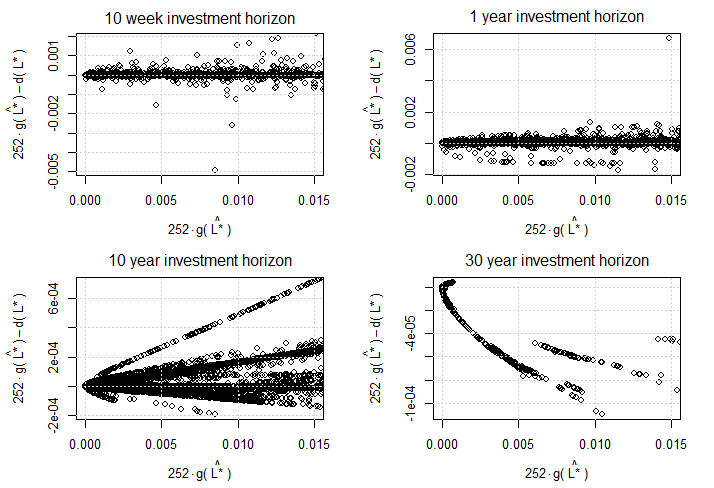}
  \caption{Like figure \ref{fig:dg}, but constricted to $252\cdot g(\hat{L^*})\leq .015$.}
  \label{fig:dgminus2}
\end{figure}

\begin{figure}
  \includegraphics[width=\linewidth]{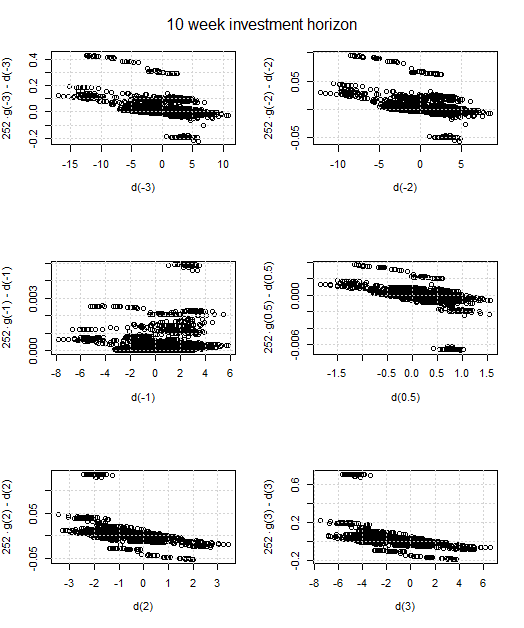}
  \caption{Using the S\&P 500 data, illustrates the error between $d(L)$ and its estimate, $252\cdot g(L)$ (see table \ref{t:fv}). Note that $n=10\cdot 5$, assuming 5 trading days per week. Each point represents a different starting date. The data spans December 29, 1927 to September 29, 2023, so starting dates are from December 29, 1927 to 10 weeks before September 29, 2023.}
  \label{fig:10wkArb}
\end{figure}

\begin{figure}
  \includegraphics[width=\linewidth]{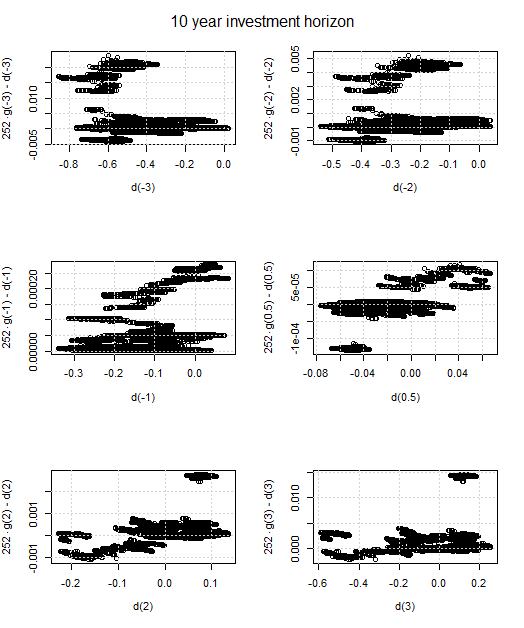}
  \caption{Same as figure \ref{fig:10wkArb}, but with $n=10\cdot 252$.}
  \label{fig:10yrArb}
\end{figure}

\begin{figure}
  \includegraphics[width=\linewidth]{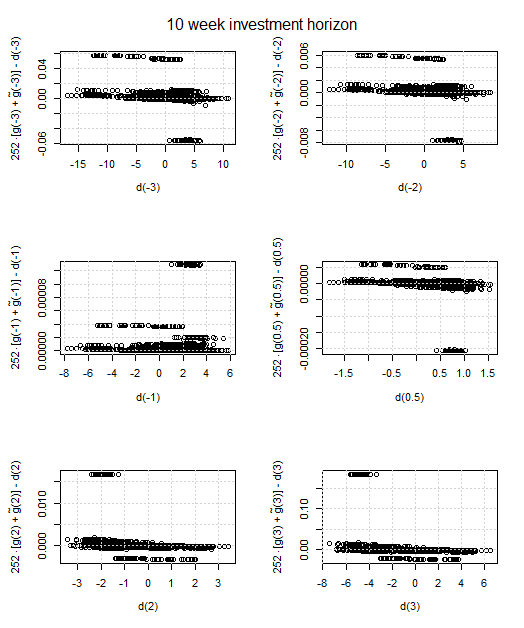}
  \caption{Using the S\&P 500 data, illustrates the error between $d(L)$ and its estimate, $252\cdot [g(L)+\widetilde{g}(L)]$ (see table \ref{t:fv}). Note that $n=10\cdot 5$, assuming 5 trading days per week. Each point represents a different starting date. The data spans December 29, 1927 to September 29, 2023, so starting dates are from December 29, 1927 to 10 weeks before September 29, 2023.}
  \label{fig:10wkArb34}
\end{figure}

\begin{figure}
  \includegraphics[width=\linewidth]{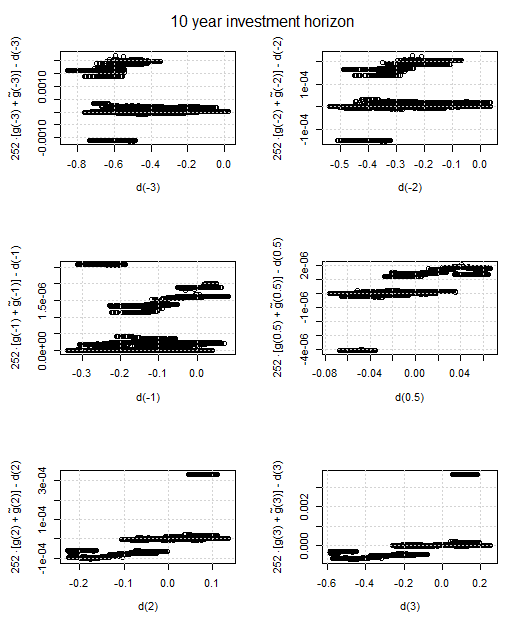}
  \caption{Same as figure \ref{fig:10wkArb34}, but with $n=10\cdot 252$.}
  \label{fig:10yrArb34}
\end{figure}

\begin{figure}
  \includegraphics[width=\linewidth]{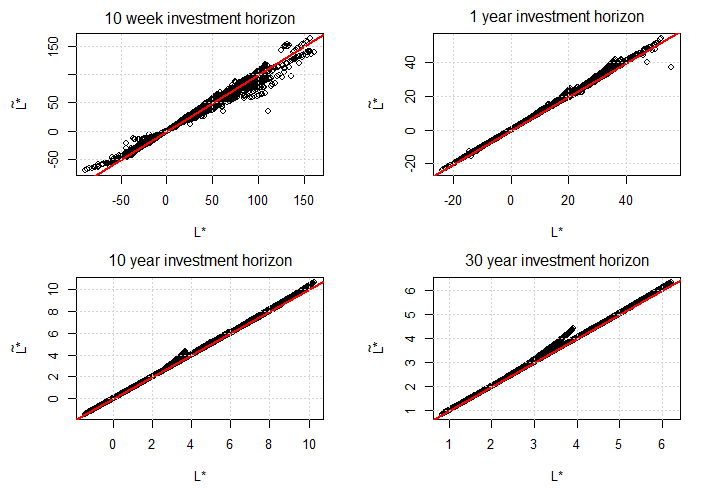}
  \caption{Analogue of figure \ref{fig:llhat}, but for $\tilde{L^*}$ instead of $\hat{L^*}$.}
  \label{fig:lltilde}
\end{figure}

\begin{figure}
  \includegraphics[width=\linewidth]{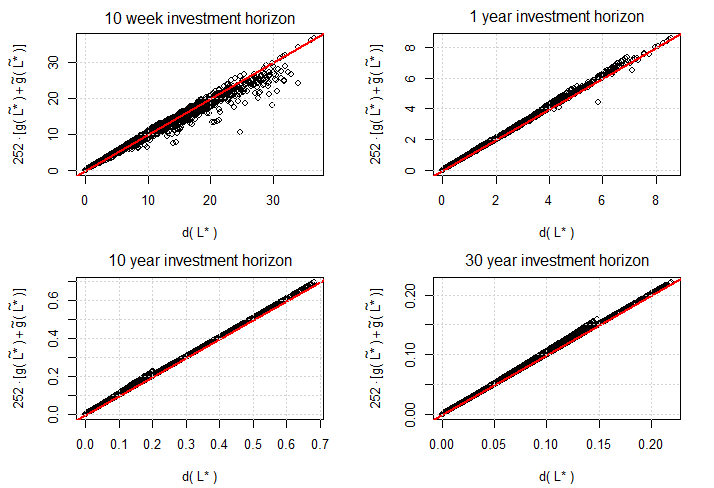}
  \caption{Analogue of figure \ref{fig:llhat}, but for $252\cdot [g(\tilde{L^*})+\tilde{g}(\tilde{L^*})]$ instead of $252\cdot g(\hat{L^*})$.}
  \label{fig:dg34}
\end{figure}

\begin{figure}
  \includegraphics[width=\linewidth]{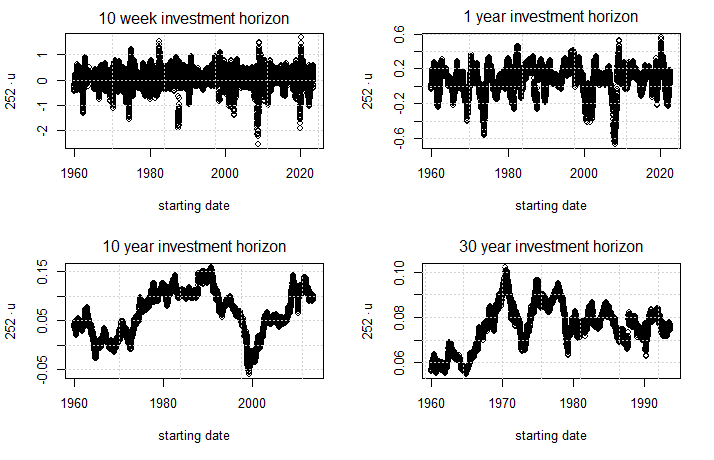}
  \caption{Illustrates $u$ for various starting dates and investment horizons. Note that $n=10\cdot 5,\ 252,\ 10\cdot 252,\ 30\cdot 252$ for 10 week, 1 year, 10 year and 30 year investment horizons. The data spans January 4, 1960 to September 29, 2023. For example, the 10 week investment horizon has starting dates from January 4, 1960 to 10 weeks before September 29, 2023.}
  \label{fig:u}
\end{figure}

\begin{figure}
  \includegraphics[width=\linewidth]{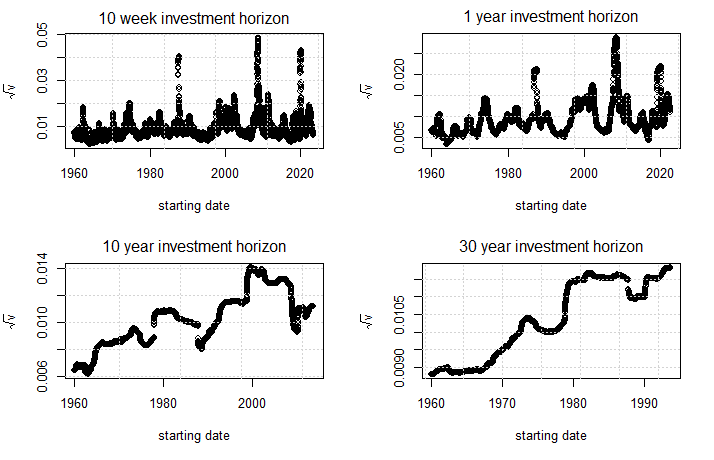}
  \caption{Analogue of figure \ref{fig:u}, but for $v$ instead of $u$.}
  \label{fig:v}
\end{figure}

\begin{figure}
  \includegraphics[width=\linewidth]{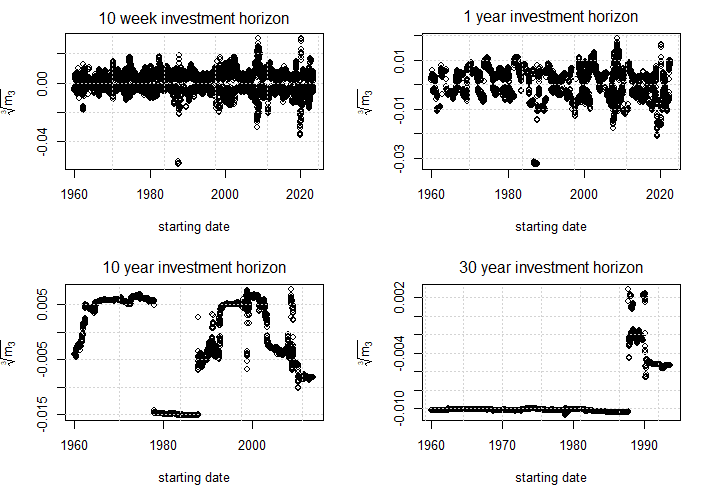}
  \caption{Analogue of figure \ref{fig:u}, but for $m_3$ instead of $u$.}
  \label{fig:m3}
\end{figure}

\begin{figure}
  \includegraphics[width=\linewidth]{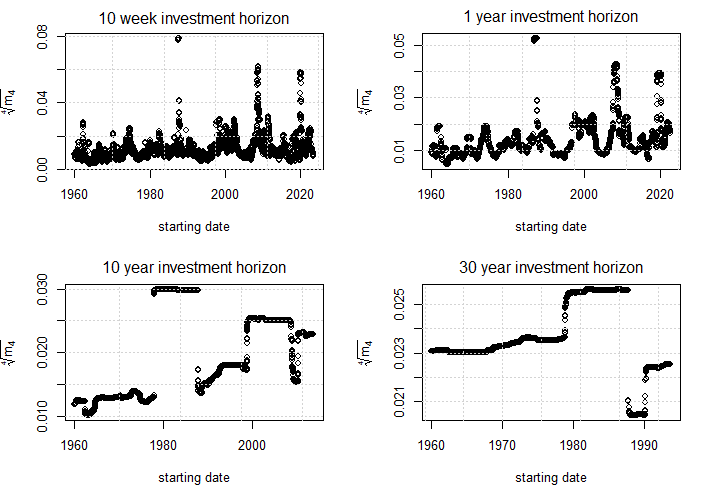}
  \caption{Analogue of figure \ref{fig:u}, but for $m_4$ instead of $u$.}
  \label{fig:m4}
\end{figure}

\begin{figure}
  \includegraphics[width=\linewidth]{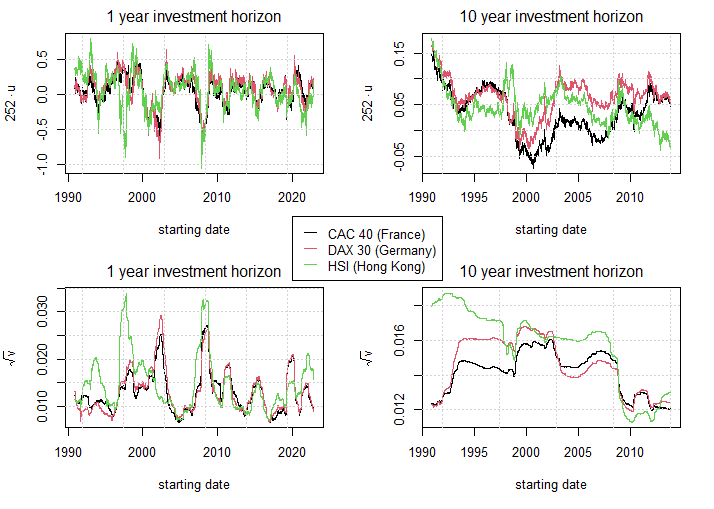}
  \caption{Analogue of figures \ref{fig:u} and \ref{fig:v}, but for foreign indexes.}
  \label{fig:fghs1}
\end{figure}

\begin{figure}
  \includegraphics[width=\linewidth]{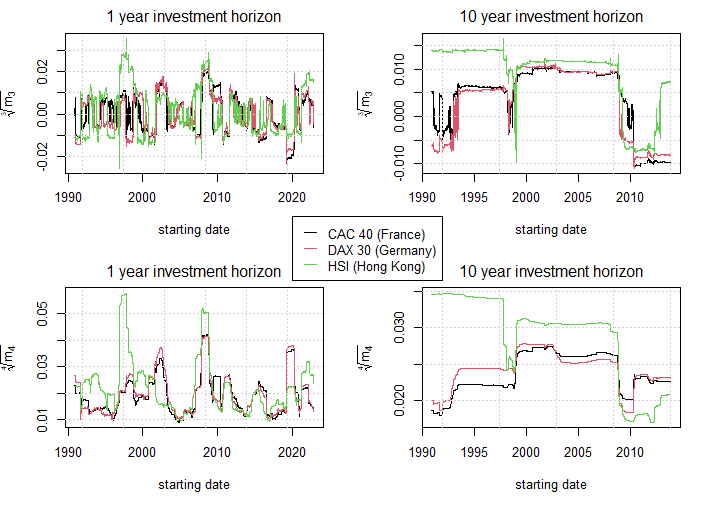}
  \caption{Analogue of figures \ref{fig:m3} and \ref{fig:m4}, but for foreign indexes.}
  \label{fig:fghs2}
\end{figure}

\begin{table}
\begin{center}
\caption{For various annualized mean daily log-returns of the underlying benchmark index ($252u$) and daily volatilities $(\sqrt{v})$, the gray columns show $252\cdot g(L)$ for $L=-3$. Additionally, 
\begin{equation*}
252\cdot g(L)-[\eqref{eq:min}^*-252\cdot(\delta_1+\delta_5)]\ \text{and}\ \eqref{eq:max}^*+252\cdot(\delta_1+\delta_5)-252\cdot g(L)
\end{equation*}
are given to the left and right of the gray columns, respectively. Note that $\eqref{eq:min}^*$ and $\eqref{eq:max}^*$ were computed using $\overline{z}=-\underline{z}=.25$, $\boldsymbol\delta$ as in \eqref{eq:deltachoice}, $m_3\in[\underline{m_3},\overline{m_3}]$, $m_4\in[\underline{m_4},\overline{m_4}]$ and $\underline{m_3},\overline{m_3},\underline{m_4},\overline{m_4}$ as in \eqref{eq:m3m4app}.}\label{t:n3}
\begin{tabular}{ c|csc|csc|csc } 
 \bottomrule
 $\sqrt{v}$ & \multicolumn{3}{|c|}{$252u=-.2$} & \multicolumn{3}{|c|}{$252u=-.08$} & \multicolumn{3}{|c}{$252u=-.02$} \\ \hline
0.005& 0.036& 0.762& 0.008& 0.037& 0.282& 0.008& 0.037& 0.042& 0.008\\
0.01& 0.053& 0.649& 0.015& 0.053& 0.169& 0.015& 0.053& -0.071& 0.015\\
0.015& 0.053& 0.46& 0.016& 0.053& -0.02& 0.016& 0.053& -0.26& 0.016\\
0.02& 0.052& 0.195& 0.016& 0.052& -0.285& 0.016& 0.052& -0.525& 0.016\\
0.025& 0.05& -0.145& 0.015& 0.05& -0.625& 0.015& 0.05& -0.865& 0.015\\
0.03& 0.047& -0.561& 0.013& 0.047& -1.041& 0.013& 0.047& -1.281& 0.013\\
 \toprule
\end{tabular}
\vspace{\baselineskip}
\begin{tabular}{ c|csc|csc|csc } 
 \bottomrule
 $\sqrt{v}$ & \multicolumn{3}{|c|}{$252u=.02$} & \multicolumn{3}{|c|}{$252u=.08$} & \multicolumn{3}{|c}{$252u=.2$} \\ \hline
0.005& 0.037& -0.118& 0.008& 0.037& -0.358& 0.008& 0.037& -0.838& 0.008\\
0.01& 0.053& -0.231& 0.015& 0.053& -0.471& 0.015& 0.053& -0.951& 0.015\\
0.015& 0.053& -0.42& 0.016& 0.053& -0.66& 0.016& 0.053& -1.14& 0.016\\
0.02& 0.052& -0.685& 0.016& 0.052& -0.925& 0.016& 0.052& -1.405& 0.016\\
0.025& 0.05& -1.025& 0.015& 0.05& -1.265& 0.015& 0.05& -1.745& 0.015\\
0.03& 0.047& -1.441& 0.013& 0.047& -1.681& 0.013& 0.047& -2.161& 0.013\\
 \toprule
\end{tabular}
\end{center}
\end{table}

\begin{table}
\begin{center}
\caption{Same as table \ref{t:n3}, but with $L=-2$.}\label{t:n2}
\begin{tabular}{ c|csc|csc|csc } 
 \bottomrule
 $\sqrt{v}$ & \multicolumn{3}{|c|}{$252u=-.2$} & \multicolumn{3}{|c|}{$252u=-.08$} & \multicolumn{3}{|c}{$252u=-.02$} \\ \hline
0.005& 0.007& 0.581& 0.002& 0.007& 0.221& 0.002& 0.007& 0.041& 0.002\\
0.01& 0.009& 0.524& 0.004& 0.009& 0.164& 0.004& 0.009& -0.016& 0.004\\
0.015& 0.009& 0.43& 0.004& 0.009& 0.07& 0.004& 0.009& -0.11& 0.004\\
0.02& 0.009& 0.298& 0.004& 0.009& -0.062& 0.004& 0.009& -0.242& 0.004\\
0.025& 0.009& 0.127& 0.004& 0.009& -0.233& 0.004& 0.009& -0.413& 0.004\\
0.03& 0.009& -0.08& 0.004& 0.009& -0.44& 0.004& 0.009& -0.62& 0.004\\
 \toprule
\end{tabular}
\vspace{\baselineskip}
\begin{tabular}{ c|csc|csc|csc } 
 \bottomrule
 $\sqrt{v}$ & \multicolumn{3}{|c|}{$252u=.02$} & \multicolumn{3}{|c|}{$252u=.08$} & \multicolumn{3}{|c}{$252u=.2$} \\ \hline
0.005& 0.007& -0.079& 0.002& 0.007& -0.259& 0.002& 0.007& -0.619& 0.002\\
0.01& 0.009& -0.136& 0.004& 0.009& -0.316& 0.004& 0.009& -0.676& 0.004\\
0.015& 0.009& -0.23& 0.004& 0.009& -0.41& 0.004& 0.009& -0.77& 0.004\\
0.02& 0.009& -0.362& 0.004& 0.009& -0.542& 0.004& 0.009& -0.902& 0.004\\
0.025& 0.009& -0.533& 0.004& 0.009& -0.713& 0.004& 0.009& -1.072& 0.004\\
0.03& 0.009& -0.74& 0.004& 0.009& -0.92& 0.004& 0.009& -1.28& 0.004\\
 \toprule
\end{tabular}
\end{center}
\end{table}

\begin{table}
\begin{center}
\caption{Same as table \ref{t:n3}, but with $L=-1$.}\label{t:n1}
\begin{tabular}{ c|csc|csc|csc } 
 \bottomrule
 $\sqrt{v}$ & \multicolumn{3}{|c|}{$252u=-.2$} & \multicolumn{3}{|c|}{$252u=-.08$} & \multicolumn{3}{|c}{$252u=-.02$} \\ \hline
0.005& 0& 0.394& 0& 0& 0.154& 0& 0& 0.034& 0\\
0.01& 0.001& 0.375& 0& 0.001& 0.135& 0& 0.001& 0.015& 0\\
0.015& 0.001& 0.343& 0& 0.001& 0.103& 0& 0.001& -0.017& 0\\
0.02& 0.001& 0.299& 0& 0.001& 0.059& 0& 0.001& -0.061& 0\\
0.025& 0.001& 0.242& 0& 0.001& 0.002& 0& 0.001& -0.118& 0\\
0.03& 0.001& 0.173& 0& 0.001& -0.067& 0& 0.001& -0.187& 0\\
 \toprule
\end{tabular}
\vspace{\baselineskip}
\begin{tabular}{ c|csc|csc|csc } 
 \bottomrule
 $\sqrt{v}$ & \multicolumn{3}{|c|}{$252u=.02$} & \multicolumn{3}{|c|}{$252u=.08$} & \multicolumn{3}{|c}{$252u=.2$} \\ \hline
0.005& 0& -0.046& 0& 0& -0.166& 0& 0& -0.406& 0\\
0.01& 0.001& -0.065& 0& 0.001& -0.185& 0& 0.001& -0.425& 0\\
0.015& 0.001& -0.097& 0& 0.001& -0.217& 0& 0.001& -0.457& 0\\
0.02& 0.001& -0.141& 0& 0.001& -0.261& 0& 0.001& -0.501& 0\\
0.025& 0.001& -0.198& 0& 0.001& -0.318& 0& 0.001& -0.557& 0\\
0.03& 0.001& -0.267& 0& 0.001& -0.387& 0& 0.001& -0.627& 0\\
 \toprule
\end{tabular}
\end{center}
\end{table}

\begin{table}
\begin{center}
\caption{Same as table \ref{t:n3}, but with $L=.5$.}\label{t:p5}
\begin{tabular}{ c|csc|csc|csc } 
 \bottomrule
 $\sqrt{v}$ & \multicolumn{3}{|c|}{$252u=-.2$} & \multicolumn{3}{|c|}{$252u=-.08$} & \multicolumn{3}{|c}{$252u=-.02$} \\ \hline
0.005& 0& 0.101& 0& 0& 0.041& 0& 0& 0.011& 0\\
0.01& 0& 0.103& 0& 0& 0.043& 0& 0& 0.013& 0\\
0.015& 0& 0.107& 0& 0& 0.047& 0& 0& 0.017& 0\\
0.02& 0& 0.113& 0& 0& 0.053& 0& 0& 0.023& 0\\
0.025& 0& 0.12& 0& 0& 0.06& 0& 0& 0.03& 0\\
0.03& 0& 0.128& 0& 0& 0.068& 0& 0& 0.038& 0\\
 \toprule
\end{tabular}
\vspace{\baselineskip}
\begin{tabular}{ c|csc|csc|csc } 
 \bottomrule
 $\sqrt{v}$ & \multicolumn{3}{|c|}{$252u=.02$} & \multicolumn{3}{|c|}{$252u=.08$} & \multicolumn{3}{|c}{$252u=.2$} \\ \hline
0.005& 0& -0.009& 0& 0& -0.039& 0& 0& -0.099& 0\\
0.01& 0& -0.007& 0& 0& -0.037& 0& 0& -0.097& 0\\
0.015& 0& -0.003& 0& 0& -0.033& 0& 0& -0.093& 0\\
0.02& 0& 0.003& 0& 0& -0.027& 0& 0& -0.087& 0\\
0.025& 0& 0.01& 0& 0& -0.02& 0& 0& -0.08& 0\\
0.03& 0& 0.018& 0& 0& -0.012& 0& 0& -0.072& 0\\
 \toprule
\end{tabular}
\end{center}
\end{table}

\begin{table}
\begin{center}
\caption{Same as table \ref{t:n3}, but with $L=2$.}\label{t:2}
\begin{tabular}{ c|csc|csc|csc } 
 \bottomrule
 $\sqrt{v}$ & \multicolumn{3}{|c|}{$252u=-.2$} & \multicolumn{3}{|c|}{$252u=-.08$} & \multicolumn{3}{|c}{$252u=-.02$} \\ \hline
0.005& 0.006& -0.206& 0.002& 0.006& -0.086& 0.002& 0.006& -0.026& 0.002\\
0.01& 0.008& -0.225& 0.004& 0.008& -0.105& 0.004& 0.008& -0.045& 0.004\\
0.015& 0.008& -0.257& 0.004& 0.008& -0.137& 0.004& 0.008& -0.077& 0.004\\
0.02& 0.008& -0.301& 0.004& 0.008& -0.181& 0.004& 0.008& -0.121& 0.004\\
0.025& 0.008& -0.358& 0.004& 0.008& -0.238& 0.004& 0.008& -0.178& 0.004\\
0.03& 0.008& -0.427& 0.004& 0.008& -0.307& 0.004& 0.008& -0.247& 0.004\\
 \toprule
\end{tabular}
\vspace{\baselineskip}
\begin{tabular}{ c|csc|csc|csc } 
 \bottomrule
 $\sqrt{v}$ & \multicolumn{3}{|c|}{$252u=.02$} & \multicolumn{3}{|c|}{$252u=.08$} & \multicolumn{3}{|c}{$252u=.2$} \\ \hline
0.005& 0.006& 0.014& 0.002& 0.006& 0.074& 0.002& 0.006& 0.194& 0.002\\
0.01& 0.008& -0.005& 0.004& 0.008& 0.055& 0.004& 0.008& 0.175& 0.004\\
0.015& 0.008& -0.037& 0.004& 0.008& 0.023& 0.004& 0.008& 0.143& 0.004\\
0.02& 0.008& -0.081& 0.004& 0.008& -0.021& 0.004& 0.008& 0.099& 0.004\\
0.025& 0.008& -0.138& 0.004& 0.008& -0.078& 0.004& 0.008& 0.042& 0.004\\
0.03& 0.008& -0.207& 0.004& 0.008& -0.147& 0.004& 0.008& -0.027& 0.004\\
 \toprule
\end{tabular}
\end{center}
\end{table}

\begin{table}
\begin{center}
\caption{Same as table \ref{t:n3}, but with $L=3$.}\label{t:3}
\begin{tabular}{ c|csc|csc|csc } 
 \bottomrule
 $\sqrt{v}$ & \multicolumn{3}{|c|}{$252u=-.2$} & \multicolumn{3}{|c|}{$252u=-.08$} & \multicolumn{3}{|c}{$252u=-.02$} \\ \hline
0.005& 0.035& -0.419& 0.008& 0.036& -0.179& 0.008& 0.036& -0.059& 0.008\\
0.01& 0.051& -0.476& 0.014& 0.051& -0.236& 0.014& 0.051& -0.116& 0.014\\
0.015& 0.051& -0.57& 0.015& 0.051& -0.33& 0.015& 0.051& -0.21& 0.015\\
0.02& 0.05& -0.702& 0.015& 0.051& -0.462& 0.015& 0.051& -0.342& 0.015\\
0.025& 0.049& -0.873& 0.015& 0.049& -0.633& 0.015& 0.049& -0.513& 0.015\\
0.03& 0.045& -1.08& 0.013& 0.045& -0.84& 0.013& 0.045& -0.72& 0.013\\
 \toprule
\end{tabular}
\vspace{\baselineskip}
\begin{tabular}{ c|csc|csc|csc } 
 \bottomrule
 $\sqrt{v}$ & \multicolumn{3}{|c|}{$252u=.02$} & \multicolumn{3}{|c|}{$252u=.08$} & \multicolumn{3}{|c}{$252u=.2$} \\ \hline
0.005& 0.036& 0.021& 0.008& 0.036& 0.141& 0.008& 0.035& 0.381& 0.008\\
0.01& 0.051& -0.036& 0.014& 0.051& 0.084& 0.014& 0.051& 0.324& 0.014\\
0.015& 0.051& -0.13& 0.015& 0.051& -0.01& 0.015& 0.051& 0.23& 0.015\\
0.02& 0.051& -0.262& 0.015& 0.051& -0.142& 0.015& 0.051& 0.098& 0.016\\
0.025& 0.049& -0.433& 0.015& 0.049& -0.313& 0.015& 0.049& -0.073& 0.015\\
0.03& 0.045& -0.64& 0.013& 0.045& -0.52& 0.013& 0.045& -0.28& 0.013\\
 \toprule
\end{tabular}
\end{center}
\end{table}

\begin{algorithm}
\caption{Computes $m$ and $z_1,z_2,...,z_m$}
\label{alg:zi}
\begin{algorithmic}
\Require $\underline{z}<0<\overline{z}$ 
\Require $\boldsymbol\delta=(\delta_1,\delta_2,\delta_3,\delta_4,\delta_5)^T>\mathbf{0}$ \Comment{Error bounds}%$B=(10^{-6},\ 10^{-8},\ 10^{-10},\ 10^{-5}/252,\ 10^{-5}/252)^T$ 
\Require \begin{equation*}F(x,\delta)=\begin{pmatrix}\vert\psi_{x,\delta}(t^*)\vert_{\phi(x)=\log(1+x)}\\\vert\psi_{x,\delta}(t^*)\vert_{\phi(x)=x^2}\\\vert\psi_{x,\delta}(t^*)\vert_{\phi(x)=x^3}\\\vert\psi_{x,\delta}(t^*)\vert_{\phi(x)=x^4}\\\vert\psi_{x,\delta}(t^*)\vert_{\phi(x)=\log(1+Lx)})\end{pmatrix}\end{equation*}\Comment{See section \ref{a:bounds}}
\State $j\gets 1$
\State $z_j\gets \underline{z}$
\While{$z_j<0$}
\If{$F(z_j,0-z_j)\leq \boldsymbol\delta$} \Comment{Check the inequality element-wise}
\State $z_{j+1}\gets 0$
\State $j\gets j+1$
\State \textbf{break}
\EndIf
\State $k\gets \min\{k_0:\ k_0\in\{2,2.1,2.2,...\},\ z_j+10^{-k_0}<0\}$
\State $\delta\gets10^{-k}$
\While{$! (F(z_j,\delta)\leq \boldsymbol\delta)$} \Comment{While the inequality does not hold}
\State $k\gets k+.1$
\State $\delta\gets10^{-k}$
\EndWhile
\State $z_{j+1}\gets z_j+\delta$
\State $j\gets j+1$
\EndWhile
\While{$z_j<\overline{z}$}
\If{$F(z_j,\overline{z}-z_j)\leq \boldsymbol\delta$} 
\State $z_{j+1}\gets \overline{z}$
\State $j\gets j+1$
\State \textbf{break}
\EndIf
\State $k\gets \min\{k_0:\ k_0\in\{2,2.1,2.2,...\},\ z_j+10^{-k_0}<\overline{z}\}$
\State $\delta\gets10^{-k}$
\While{$! (F(z_j,\delta)\leq \boldsymbol\delta)$} 
\State $k\gets k+.1$
\State $\delta\gets10^{-k}$
\EndWhile
\State $z_{j+1}\gets z_j+\delta$
\State $j\gets j+1$
\EndWhile
\State $m\gets j$\\
\Return{$m$ and $z_1,z_2,...,z_m$}
\end{algorithmic}
\end{algorithm}

\end{document}